\UseRawInputEncoding
\documentclass[aps, prx, twocolumn, showpacs, floatfix,10pt,superscriptaddress]{revtex4-2}
\usepackage{amsmath}
\usepackage{amsfonts}
\usepackage{amsbsy}
\usepackage{amssymb}
\usepackage{graphicx}
\usepackage{textcomp}
\usepackage[caption=false,singlelinecheck=false]{subfig}
\usepackage{xcolor}
\usepackage{mathrsfs}
\usepackage{mathtools}
\usepackage{bm}
\usepackage{gensymb}
\usepackage{braket,wasysym}
\usepackage{float}
\usepackage{tikz}
\usepackage[colorlinks,linkcolor=blue,citecolor=red,filecolor=magenta,urlcolor=red,breaklinks]{hyperref}
\usepackage[bottom]{footmisc}
\usepackage{verbatim}

\hypersetup{colorlinks=true, urlcolor=magenta, citecolor=red, pdfborder={0 0 0}}
\usepackage{breakurl}
\usepackage{natbib}

\allowdisplaybreaks

\normalfont
\def\be{\begin{equation}}
	\def\ee{\end{equation}}
\def\bea{\begin{eqnarray}}
	\def\eea{\end{eqnarray}}

\begin{document}
\title{
Non-Fermi-Liquid Behaviors in Weakly Interacting Two-Dimensional Quasicrystals
}

\author{Junmo Jeon}
\email{junmojeon@sophia.ac.jp}
\author{Shiro Sakai}
\email{shirosakai@sophia.ac.jp}
\affiliation{Physics Division, Sophia University, Chiyoda-ku, Tokyo 102-8554, Japan}

\date{\today}
\begin{abstract}
Non-Fermi-liquid (NFL) behavior is usually associated with strong electronic correlations, quantum criticality, or singular fluctuations that invalidate the quasiparticle picture. Quasicrystals are known to provide a fundamentally different route to such physics: their lack of translational symmetry gives rise to critical single-particle states and fragmented spectra even in the absence of electron-electron interactions. Here, we numerically study 
the NFL behaviors in quasicrystals with 
weak interactions. Using Penrose and Ammann--Beenker tilings as representative quasiperiodic systems, we analyze thermodynamic, magnetic, and transport properties within a weak-coupling perturbative framework. The Sommerfeld coefficient and magnetic susceptibility show anomalous temperature dependences governed by the singular energy dependence of the density of states near the Fermi level. The Wilson ratio 
reveals that low-temperature NFL behavior in quasicrystals is 
accompanied by enhanced spin fluctuations over a broad range of filling fractions, while 
its magnitude and 
temperature dependence 
vary sensitively with filling fraction. The self-energy shows anomalous scaling and a finite residual scattering rate, indicating the breakdown of 
quasiparticles.
Our site-resolved analysis reveals 
a pronounced enhancement of the residual scattering rate in regions with a high local density of states. 
These 
results serve as a guide to analyze experimental data for quasicrystals
and to identify the microscopic mechanism
of the NFL behaviors.
\end{abstract}
\maketitle

\section{Introduction}
Landau Fermi-liquid theory provides the standard framework for understanding weakly interacting metallic systems~\cite{phillips2012advanced,landau1956fermi}. Its central assumption is that the low-energy excitations of an interacting electron system can be adiabatically connected to those of a non-interacting metallic Fermi gas, resulting in well-defined quasiparticle states with a 
long lifetime~\cite{baym2008landau}.
This paradigm successfully describes a broad range of crystalline metals and serves as the basis for analyzing numerous experimental observations~\cite{baym2008landau}. However, there are many quantum materials that exhibit clear deviations from this paradigm, commonly referred to as non-Fermi-liquid (NFL) behavior~\cite{Stewart2001,lohneysen2007fermi,gegenwart2008quantum,lee2006doping,sachdev1999quantum,hartnoll2022planckian}. Such behavior is often attributed to strong electron correlations, or proximity to a quantum critical point~\cite{sachdev1999quantum,gegenwart2008quantum,keimer2015quantum,lee2006doping,hertz1976quantum,millis1993effect,hartnoll2022planckian}.

NFL behavior has been observed in a wide variety of systems, ranging from heavy-fermion compounds, 
pseudogap state of high-temperature superconducting cuprates,
and quantum critical metals~\cite{Stewart2001,keimer2015quantum,lohneysen2007fermi,gegenwart2008quantum,lee2006doping,sachdev1999quantum,hertz1976quantum,millis1993effect,Timusk1999,deguchi2012quantum,Andrade2015,Sato2022effects,hartnoll2022planckian}. Owing to unconventional thermodynamic, magnetic, and transport properties distinct from those of Landau Fermi liquids, various NFL systems and their physical properties have emerged as a central topic of interest among researchers over the years~\cite{lohneysen2007fermi,Stewart2001}. In conventional crystalline systems, such deviations from Fermi-liquid behavior are generally understood as consequences of
strong electron-electron correlations that destabilize the Landau quasiparticle picture~\cite{si2010heavy,coleman2005quantum,sachdev1999quantum}. 
In fact, in the weakly interacting regime, Fermi-liquid theory has been quite successful in analyzing the experimental data measured for periodic materials~\cite{baym2008landau,leggett1975theoretical,pines1967theory}.
In contrast, for aperiodic systems where conventional wave numbers cannot characterize quasiparticle states, the applicability of Fermi-liquid theory itself remains unclear even in the weak-coupling regime~\cite{janot2012quasicrystals,kellendonk2015mathematics,baake2013aperiodic}.

The situation is much less clear in quasicrystals, which are long-range ordered without periodicity~\cite{shechtman1984metallic,jagannathan2021fibonacci,levine1984quasicrystals,janot2012quasicrystals}.
Theoretical studies suggest that single-electron wavefunctions in quasicrystals are critical, 
neither fully extended nor localized. These electronic states exhibit power-law scaling and multifractal structures in real space~\cite{kohmoto1983localization,niu1986renormalization,kohmoto1987critical,Tokihiro1988,mace2017critical,hiramoto1992electronic}. Experiments on quasicrystals have also shown 
that their physical properties, 
e.g., 
transport characteristics, are distinct from those of both ordinary crystalline metals~\cite{stadnik1998physical,roche1998fermi,fujiwara1994electronic,roche1997conductivity,roche1997electronic,trambly2006quantum,trambly2013anomalous,kimura1989electronic,deguchi2012quantum,Dolinsek2012,PhysRevB.59.308,Sato2022effects}.
Thus, the applicability of 
Fermi-liquid theory to 
quasicrystals
is questionable, even for weak electron-electron interactions. Nevertheless, since a comprehensive phenomenological framework for describing electronic states in weakly interacting quasicrystals is still lacking,
many experimental results have been analyzed with the Fermi-liquid theory, which may lead an incorrect conclusion for these systems~\cite{jagannathan2021fibonacci,Dolinsek2012,roche1997electronic,PhysRevB.59.308,Poon1992}. Therefore, it is highly desired to clarify the metallic behaviors of weakly-interacting quasicrystals and construct a phenomenology applicable to these materials.
In particular, the experimentally observed residual resistivity at low temperatures, remain poorly understood from a theoretical perspective~\cite{,roche1997electronic,kimura1989electronic}. This raises a fundamental question: what is the 
nature of the weakly interacting electronic state in quasicrystals? More importantly, if quasiperiodicity invalidates the conventional Fermi-liquid description, how does NFL behavior emerge, and what underlying principles determine its low-energy phenomenology?

In this work, we address these questions by studying weakly interacting electron systems on two-dimensional (2D) Penrose~\cite{Penrose1974} and Ammann--Beenker tilings~\cite{Beenker1982,ammann1992aperiodic}. A Hartree-dressed 
perturbation theory
is employed to investigate thermodynamic, magnetic, and transport properties. NFL behavior 
emerge generically over broad filling ranges. We 
characterize these NFL electron states through their specific heat, magnetic susceptibility, Wilson ratio~\cite{Stewart2001,Hewson1993,coleman2005quantum}, and self-energy~\cite{Abrikosov1975} to elucidate the form of NFL behavior that arises in quasicrystals.

Importantly, these NFL signatures are strongly influenced by the unique structure of the (local) density of states [(L)DOS] of quasicrystals. The anomalous temperature dependences of the specific heat and magnetic susceptibility are controlled by the DOS near the Fermi energy. Furthermore, we show that low-temperature NFL behavior in quasicrystals is generally accompanied by enhanced spin fluctuations over a broad range of filling fractions, while both the magnitude and the temperature dependence of the Wilson ratio vary sensitively with filling fraction, reflecting the corresponding low-energy DOS near the Fermi level. Notably, the Wilson ratio remained well above unity over a wide range of filling fractions irrespective of interactions, indicating that the enhanced magnetic fluctuations driving the NFL behavior are intrinsic to the quasicrystalline electronic structure and DOS. Meanwhile, the anomalous self-energy scaling and finite residual scattering rate related to the experimental residual resistivity originate from spatially inhomogeneous LDOS. In particular, we find that the local residual scattering rate scales with the cube of the LDOS. Our work establishes a framework for the weakly interacting electron state of quasicrystals, a regime for which no counterpart to conventional Fermi-liquid phenomenology currently exists. The results presented here provide a direct link between filling fractions and experimentally observable NFL behavior, paving the way toward a systematic description of correlated metallic quasicrystals.

The remainder of this paper is organized as follows. In Sec.~\ref{sec:modelmethod}, we introduce the model and methodology. In Sec.~\ref{sec:results}, we demonstrate the NFL thermodynamic, magnetic, and transport properties of weakly interacting quasicrystals, and establish their connection to the fragmented spectrum and multifractal critical states characteristic of quasiperiodic systems. In Sec.~\ref{sec:discussion}, we discuss the broader implications of these findings, their experimental relevance, and potential future works. Finally, Sec.~\ref{sec:conclusion} summarizes the conclusions and outlook.

\section{Model and method}
\label{sec:modelmethod}
\subsection{Weakly interacting quasicrystals}
\begin{figure}[h]
	\centering
	\includegraphics[width=0.45\textwidth]{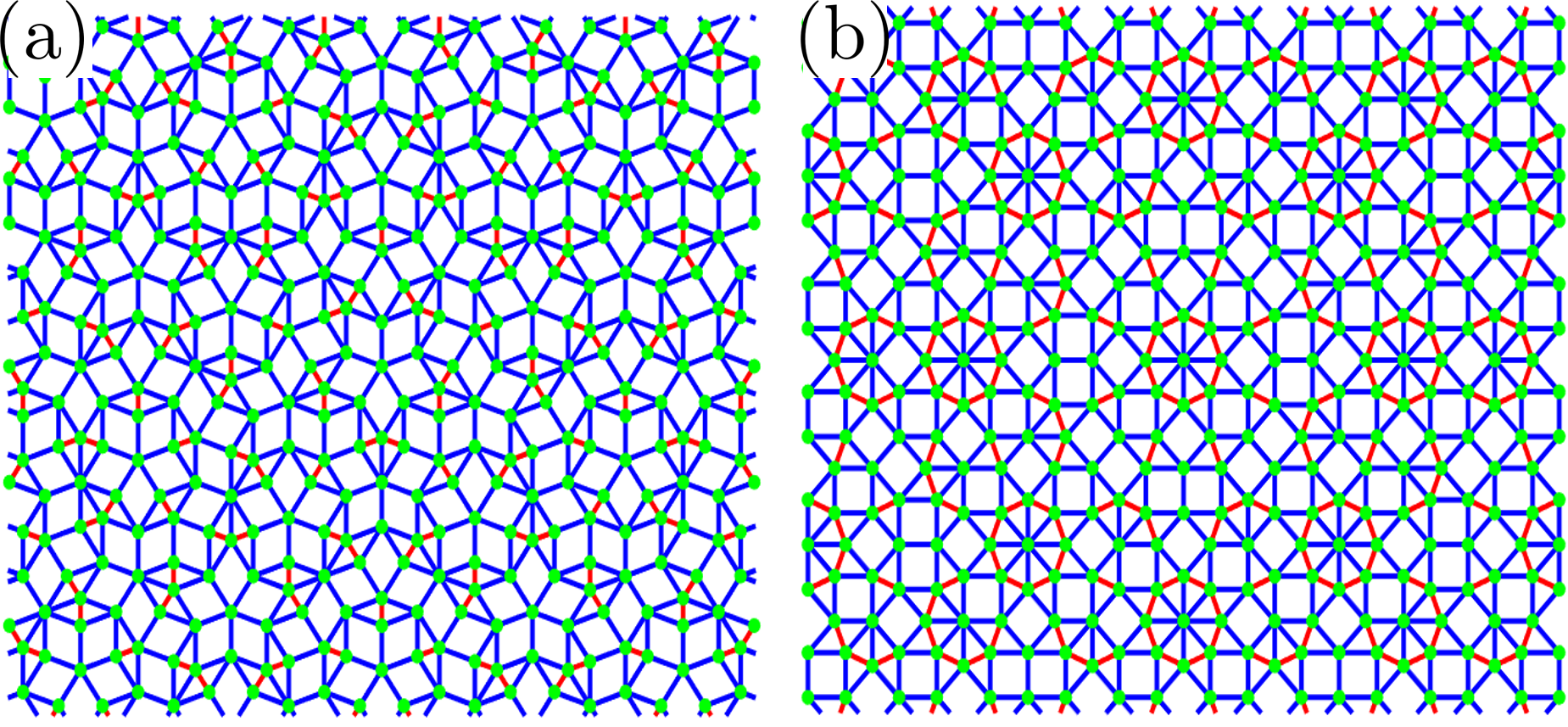}
	\caption{Tiling patterns of (a) Penrose and (b) Ammann--Beenker tilings. Green dots represent the positions of atoms. Blue and red edges represent the links with distance $a$ and $2a \sin\frac{\pi}{10}$ ($2a \sin\frac{\pi}{8}$), respectively, in the Penrose (Ammann--Beenker) tiling.}
	\label{fig:tiling}
	
\end{figure}

We consider 2D quasicrystalline lattices, focusing on the Penrose (pentagonal) \cite{Penrose1974} and Ammann--Beenker (octagonal) \cite{ammann1992aperiodic, Beenker1982} tilings as representative realizations of quasiperiodic order, i.e., long-range order without translational symmetry~\cite{kellendonk2015mathematics,janot2012quasicrystals}. Figure~\ref{fig:tiling} displays their tiling patterns. Note that both tilings do not have periodic units.
In detail, these tilings consist of a few different types of tiles, which are arranged aperiodically but rotationally symmetrically. 
Because periodicity is absent, neither the Brillouin zone nor crystal momentum is well defined in quasicrystals. Hence, the electronic states could not be described by conventional Bloch states~\cite{kohmoto1987critical,mace2017critical}.

The Penrose and Ammann--Beenker tilings can be generated through the cut-and-project scheme~\cite{kellendonk2015mathematics,baake2013aperiodic,Duneau1985}. In this construction, a subset of lattice points embedded in a higher-dimensional periodic lattice is projected onto a lower-dimensional subspace, referred to as the physical space, whose orientation is irrational with respect to the high-dimensional lattice planes. Owing to this irrational orientation, each site in the physical space is uniquely associated with a point in the orthogonal complement, known as the perpendicular space. An important property of this correspondence is that points in the perpendicular-space cluster are arranged according to the similarity of their local environments in the physical space~\cite{deBruijn1981,baake2013aperiodic}. 
Namely, physical-space sites corresponding to points within a given region of perpendicular space share a similar local environment in the physical space. 
Hence, by plotting local quantities, such as charge density, in the perpendicular space, we can see how the quantity is affected by the real-space geometry.
For example, if the quantity takes a similar value in each sectioned region of the perpendicular space, we can conclude that the quantity is mostly determined by the short-range physics. On the other hand, if the perpendicular-space map exhibits a more intricate structure, it suggests the presence of longer-range correlations, such as a fractal structure. See Appendix~\ref{A1} for details.

Let us consider single-orbital 
Hubbard 
Hamiltonian,
\begin{align}
\label{H}
&H=-\sum_{
ij,\sigma}t_{ij}\hat{c}_{i,\sigma}^\dagger\hat{c}_{j,\sigma}+U\sum_i \hat{n}_{i\uparrow}\hat{n}_{i\downarrow}-\mu\sum_{i,\sigma}\hat{n}_{i,\sigma},
\end{align}
on the quasicrystalline lattices.
Here, $\hat{c}_{i,\sigma}^\dagger (\hat{c}_{i,\sigma})$ is electronic creation (annihilation) operator at the $i$-th site with spin $\sigma=\uparrow$ or $\downarrow$. $\hat{n}_{i,\sigma}=\hat{c}_{i,\sigma}^\dagger\hat{c}_{i,\sigma}$ is the local charge density operator. $U>0$ denotes the onsite repulsive interaction strength. $\mu$ is the chemical potential. $t_{ij}$ is the hopping integral which solely depend on the distance, $r_{ij}=\vert \vec{r}_i-\vec{r}_j\vert$ where $\vec{r}_i$ is the position vector of the $i$-th site. We adopt the Slater-Koster model of the hopping integral~\cite{Slater1954,Huang2018}, $t_{ij}=t e^{-(r_{ij}-a)/a}$, where $t$ is the energy unit and $a$ is a constant. For simplicity, we assume that $t_{ij}$ is nonzero for $r_{ij}\le a$ but zero otherwise, 
with setting 
$a$ as 
the edge length of the rhombuses,
to ensure that all vertices of the tiling form the connected graph. Then, electrons can hop only between atoms separated by distances of $a$ and $2a \sin\frac{\pi}{10}$ ($2a \sin\frac{\pi}{8}$) in the Penrose (Ammann--Beenker) tiling, which are represented as blue and red edges in Fig.~\ref{fig:tiling}.

For $U=0$, the non-interacting model hosts critical 
single-electron
states in quasicrystals~\cite{Tokihiro1988,mace2017critical}. These states are neither localized nor extended but self-similar and exhibit power-law decaying wavefunction envelopes~\cite{kohmoto1987critical}, producing multifractal LDOS patterns in the Penrose and Ammann--Beenker tilings~\cite{Jeon2025,Jeon2022}.
\begin{figure}[h]
	\centering
	\includegraphics[width=0.5\textwidth]{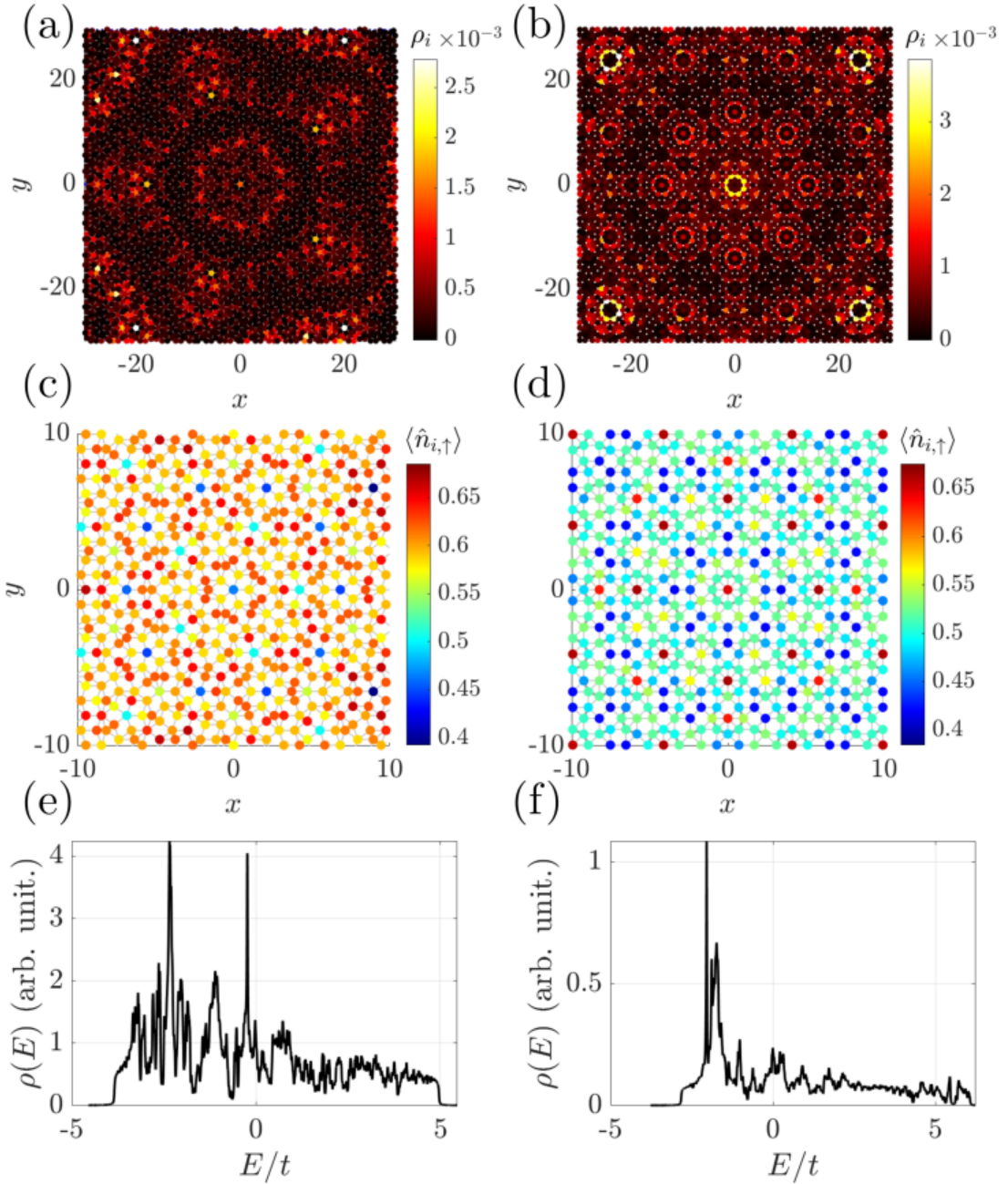}
	\caption{Real-space map of the local density of states, $\rho_i$, at the Fermi energy for (a) the Penrose tiling at 60$\%$ filling and (b) the Ammann--Beenker tiling at half filling. (c,d) The 
    local charge density for up spin, $\langle \hat{n}_{i,\uparrow}\rangle$,
    corresponding to (a) and (b), respectively.
    (e,f) Density of states corresponding to (a) and (b), respectively.
    $T=10^{-3}t$ and $U=0.5t$ are used. 
    }
	\label{fig:criticalHartree}
	
\end{figure}

Now we consider a weak repulsive interaction, $U\ll W_b$, where $W_b\sim\mathcal{O}(10t)$ is the bandwidth of the non-interacting model. 
The 
Hartree mean-field Hamiltonian is given by
\begin{align}
\label{HF}
&H_\mathrm{eff}=-\sum_{
ij,\sigma}t_{ij}\hat{c}_{i,\sigma}^\dagger\hat{c}_{j,\sigma}+U\sum_{i,\sigma} \langle\hat{n}_{i,\bar{\sigma}}\rangle\hat{n}_{i,\sigma}-\mu\sum_{i,\sigma}\hat{n}_{i,\sigma},
\end{align}
where $\bar{\sigma}$ is the opposite spin of $\sigma$. Here, 
we note that the second term, the Hartree shift, is site-dependent due to the aperiodicity of the structure.
The mean-field Hamiltonian in Eq.~\eqref{HF} can be solved self-consistently. In detail, we solve the self-consistent equation, $\langle \hat{n}_{i,\sigma}\rangle=\sum_\alpha\vert \psi^\alpha_{i,\sigma}\vert^2 f(\epsilon_\alpha,T)$ with tolerance $10^{-6}$ by updating $\mu$ to keep a given filling fraction. Here, $\psi^\alpha_{i,\sigma}$ and $\epsilon_\alpha$ are an eigenfuncion and energy of the Hamiltonian in Eq.~\eqref{HF}. $f(\epsilon_\alpha,T)=(1+e^{\epsilon_\alpha/T})^{-1}$ is the Fermi-Dirac distribution with temperature $T$. 
From here on, we simply denote it as $f_\alpha$. We use $\hbar=k_B=1$. Unless otherwise stated, we assume a paramagnetic solution and solve the self-consistent equations in the weakly interacting regime. Although this assumption would break down even for weak interactions near certain singular filling fractions, such as those at van Hove singularities, we emphasize that it does not qualitatively affect our conclusions for generic filling fractions 
in the weak-coupling regime (see Appendix~\ref{A4}).

The electronic states in quasicrystals would be multifractal critical states even in the presence of interactions \cite{hiramoto1990the}. Then, the 
LDOS
$\rho_i(E)=-\pi^{-1}\lim_{\eta\to0+}\mbox{Im}[(E+\mathrm{i}\eta-H_{\mathrm{eff}})^{-1}]_{ii}$ would also show a multifractal structure. Figures~\ref{fig:criticalHartree}(a,b) display typical multifractal 
LDOS
at the Fermi energy in the Penrose and Ammann--Beenker tilings for a small $U>0$. Consequently, the self-consistently obtained local charge field, $\langle \hat{n}_{i,\sigma}\rangle$, is inhomogeneous [see Figs.~\ref{fig:criticalHartree}(c,d)]. We emphasize that in contrast to periodic crystals, where Bloch states lead to a 
homogeneous
charge distribution, quasicrystals 
exhibit inhomogeneous charge distributions.
This inhomogeneity 
generates an aperiodic potential via the Hartree shift, further driving the electronic states of quasicrystals toward 
unconventional behavior~\cite{hiramoto1990the,sakai2021effect,sakai2022hyperuniform,Jeon2025}.

Furthermore, 
the fractal electronic states of quasicrystals affect 
the distribution of energy eigenvalues,
fragmenting the energy spectrum~\cite{Kohmoto1986,Jeon2025}.
Hence, in quasicrystals, the DOS, $\rho(E)=N^{-1}\sum_i\rho_i(E)$, where $N$ is the system size, is not smooth, which can lead to anomalous thermodynamic and electromagnetic behaviors 
even in the weak\mbox{-}interaction regime [see Figs.~\ref{fig:criticalHartree}(e,f)]. 
In this paper,
we set the system sizes to $N=7011$ and $9489$ for the Penrose and Ammann--Beenker tilings, respectively. We 
have confirmed
that our key findings are 
robust against the increase of $N$ (see Appendix~\ref{A2}).

We investigate the NFL behaviors 
arising from the interplay between critical states, inhomogeneous Hartree shifts, and the fragmented energy spectrum, in weakly interacting quasicrystals. To this end, we examine not only thermodynamic properties, 
such as
the Sommerfeld coefficient of the specific heat, but also magnetic responses such as the magnetic susceptibility and the Wilson ratio.
We further study transport properties
through an analysis of the self-energy obtained with the second-order perturbation theory.

\subsection{Sommerfeld Coefficient of the Specific Heat}
In the Fermi liquid, the low-temperature specific heat is linear in temperature, i.e., 
\begin{equation}
C(T) = \gamma T,
\end{equation}
where $\gamma$ is the Sommerfeld coefficient, proportional to the DOS
at the Fermi level~\cite{coleman2005quantum,baym2008landau}.
The specific heat, $C(T)=T \frac{dS}{dT}$, where
\begin{equation}
S(T)=-\sum_{\alpha} \left[ f_\alpha\ln f_\alpha+(1-f_\alpha)\ln(1-f_\alpha) \right]
\end{equation} 
is the entropy.
For $T \to 0$, $\gamma$ approaches a constant value, reflecting well-defined quasiparticles. In contrast, NFL behavior is characterized by deviations from this linear scaling of $C(T)$~\cite{Stewart2001,deguchi2012quantum}. To examine the temperature dependence of effective Sommerfeld coefficient, we introduce an exponent 
\begin{align}
\label{g_exponent}
g=\frac{\partial \log\gamma}{\partial\log T},
\end{align}
i.e., $\gamma\propto T^g$, where $\gamma=\gamma(T)=C(T)/T$. Note that for ideal Fermi liquid, $g=0$.

\subsection{Magnetic Susceptibility and Wilson ratio}
For the Fermi liquid, the uniform magnetic susceptibility is given by the Pauli susceptibility,
\begin{equation}
\chi(T) = \chi_0,
\end{equation}
which remains approximately constant at low temperatures, reflecting the finite 
DOS
at the Fermi level~\cite{coleman2005quantum,baym2008landau}. By contrast, NFL behavior manifests as anomalous temperature dependence of the susceptibility, e.g.,
\begin{equation}
\chi(T) \sim -\log T \quad \text{or} \quad T^{-b},
\end{equation}
with $b > 0$, indicating enhanced spin fluctuations or the absence of well-defined quasiparticles~\cite{Stewart2001,deguchi2012quantum,Andrade2015}.

To explore the magnetic susceptibility in quasicrystals, we apply a weak uniform Zeeman field, $h$, in the in-plane direction of the 2D 
quasicrystals.
This adds the spin-dependent 
onsite
potential term $h\sum_i(n_{i,\downarrow}-n_{i,\uparrow})$, which leads to the magnetization, $M(h)=\sum_{i}(\langle n_{i,\uparrow}\rangle - \langle n_{i,\downarrow}\rangle)$, and susceptibility, $\chi(T)=-\frac{\partial M}{\partial h}$. We investigate the temperature dependence of susceptibility in terms of an exponent 
\begin{align}
\label{kappa_exponent}
\kappa=\frac{\partial \log \chi}{\partial \log T},
\end{align}
i.e., $\chi\propto T^\kappa$.

To compare the magnetic response given by the susceptibility and the low-energy excitations measured through the specific heat, 
we also examine the Wilson ratio~\cite{Hewson1993},
\begin{align}
\label{Wilsonratio}
W(T)
=
\frac{4\pi^2}{3}
\frac{\chi(T)}{\gamma(T)}.
\end{align}
The Wilson ratio measures the relative enhancement of spin fluctuations compared with the low-energy thermodynamic response. For the Fermi liquid, both $\chi(T)$ and $\gamma(T)$ are proportional to the electronic 
DOS
at the Fermi energy, and hence the Wilson ratio is temperature-independent. For instance, $W(T)=1$ for free electron gas, while strongly correlated electron systems exhibit a large deviation from unity. In detail, $W>1$ indicates that magnetic fluctuations dominate the low-energy NFL behavior, as commonly observed in heavy-fermion and magnetically ordered systems~\cite{Stewart1984,vonLohneysen2007,deguchi2012quantum,Andrade2015}. In contrast, $W<1$ implies that low-energy excitations contribute more strongly to the entropy than to the magnetic response, suggesting predominantly non-magnetic excitations~\cite{Prelovsek2020,Balents2010}. In particular, a pronounced temperature dependence of $W(T)$ indicates that the magnetic response and thermodynamic excitations obey different low-energy scaling behaviors. Thus, the Wilson ratio serves as useful diagnostic of NFL behavior. 


\subsection{Self-Energy and Quasiparticle Lifetime}

In addition to thermodynamic and magnetic properties, we discuss transport properties of quasicrystals. Experimentally, quasicrystals have been known to exhibit anomalous low-temperature transport behavior, including anomalous temperature dependence and finite residual resistivity~\cite{Dolinsek2012,kimura1989electronic,Poon1992,PhysRevB.59.308,deguchi2012quantum}. From a theoretical perspective, two distinct mechanisms may contribute to such residual transport anomalies. One originates from the absence of translational symmetry, which invalidates crystal momentum as a good quantum number and leads to intrinsic elastic Bragg scattering even without electron-electron interactions~\cite{roche1997conductivity,roche1997electronic,trambly2006quantum,trambly2013anomalous,PhysRevB.43.8892}. The other arises from electron-electron interactions, which can generate a finite residual scattering rate owing to the critical nature of the underlying single-particle states~\cite{Fetter1971,Abrikosov1975}. While the former mechanism has been extensively investigated within non-interacting transport theories based on the Kubo-Greenwood formalism and quantum diffusion~\cite{roche1997conductivity,roche1998fermi,trambly2006quantum,trambly2013anomalous,PhysRevLett.70.3915}, these studies do not address how electron-electron interactions microscopically generate a quasiparticle scattering rate or how the corresponding lifetime evolves with temperature, and hence the latter remains largely unexplored. In this work, we focus on the latter mechanism and investigate its role in the residual transport anomaly by examining the quasiparticle lifetime inferred from the weak-coupling self-energy. To this end, we compute the electronic self-energy within second-order perturbation theory.

In the Fermi liquid, the imaginary part of the retarded self-energy exhibits the well-known quadratic scaling at low energies,
\begin{equation}
\mathrm{Im}\,\Sigma(\omega) \sim -\omega^2,
\end{equation}
which implies a long-lived quasiparticle with a lifetime 
$\tau \propto \omega^{-2}$
as $\omega \to 0$. By contrast, 
a NFL
behavior is characterized by anomalous scaling,
\begin{equation}
\mathrm{Im}\,\Sigma(\omega) \sim -|\omega|^\nu,
\end{equation}
with $\nu < 2$, indicating a breakdown of the quasiparticle picture~\cite{Abrikosov1975,Fetter1971}.

We begin by exploring the self-energy as a function of Matsubara frequencies. Instead of the noninteracting Green's function, we use the Hartree-dressed Green's function defined by
\begin{equation}
G_{ij}^{(H)}(i\omega_n) = \left[i\omega_n  - H_{\mathrm{eff}}\right]^{-1}_{ij},
\end{equation}
where $\omega_n=(2n+1)\pi T$ is the fermionic Matsubara frequency. This is because the Hartree shift, which is site-dependent in quasicrystals, would play 
a
crucial role 
in deforming
the single-particle wavefunctions in quasicrystals, while keeping the lifetime infinite. Here, we assume the paramagnetic solution, $G_{ij}^{(H)}=G_{ij,\uparrow}^{(H)}=G_{ij,\downarrow}^{(H)}$, as is generally expected in the weakly correlated regime, excluding the region around singularities such as the van Hove singularity. 
We will briefly discuss the possibility of non-paramagnetic solutions in Appendix \ref{A4}. The corresponding imaginary-time Green's function is given by
\begin{equation}
\label{imaginarytimeG}
G_{ij}^{(H)}(\tau) = \frac{1}{\beta} \sum_n e^{-i\omega_n \tau} G_{ij}^{(H)}(i\omega_n)
\end{equation}
with $\beta=1/T$. Note that thermal DOS is given by
\begin{align}
\label{thermaldos}
\rho(T)=-\frac{1}{N}\mbox{Tr}\left[G^{(H)}\left(\tau=\frac{1}{2T}\right)\right],
\end{align}
where $G^{(H)}(\tau)$ is the imaginary-time Green's function in Eq.~\eqref{imaginarytimeG} and $\mbox{Tr}$ stands for the trace over the site index. $\rho_i(T)=-G^{(H)}_{ii}\left(\tau=\frac{1}{2T}\right)$ defines the thermal 
LDOS.

Exploiting the locality of the onsite interaction, we assume that the higher-order correction of self-energy is site-diagonal and can be written as
\begin{equation}
\label{perturbativeselflocal}
\Sigma_i(\tau) = U^2\, G_{ii}^{(H)}(\tau)^2\, G_{ii}^{(H)}(-\tau).
\end{equation}
Note that Eq.~\eqref{perturbativeselflocal} contains not only second-order but also higher-order contributions through the Hartree term in $G^{(H)}$.
The Matsubara self-energy is obtained via Fourier transformation,
\begin{equation}
\label{higherorderselfMatsubara}
\Sigma_i(i\omega_n) = \int_0^\beta d\tau\, e^{i\omega_n \tau} \Sigma_i(\tau).
\end{equation}
For the Fermi liquid, $\mathrm{Im}\Sigma_i(\mathrm{i}\omega_n)\sim\omega_n$, while $\mathrm{Im}\Sigma_i(\mathrm{i}\omega_n)\sim\vert\omega_n\vert^{\nu_M(i)}$ with $\nu_M(i)<1$ for NFL. Note that the local self-energy given by Eq.~\eqref{higherorderselfMatsubara} 
is
inhomogeneous in quasicrystals, and hence the scaling exponent would be 
also site-dependent.

To connect with physical scattering processes, we construct an approximate eigenstate-resolved self-energy. In quasicrystals, eigenstates are not labeled by crystal momentum due to the absence of translational symmetry, and 
instead
the energy 
eigenstates provide a natural choice of the basis.
We project the local self-energy [Eq.~\eqref{higherorderselfMatsubara}] onto the eigenstates of $H_{\mathrm{eff}}$ as
\begin{equation}
\label{energybasisselfMatsubara}
\Sigma_\alpha(\mathrm{i}\omega_n)
\approx
\sum_i |\psi_\alpha(i)|^2 \, \Sigma_i(\mathrm{i}\omega_n),
\end{equation}
where $\psi_\alpha(i)$ is the eigenfunction corresponding to the eigenenergy $\epsilon_\alpha$. Note that we assume locality of the self-energy, i.e., $\Sigma_{ij}\approx\Sigma_i\delta_{ij}$.
In the following, we 
focus on the diagonal contributions in the energy-eigenstate basis. One can 
see NFL characteristics
from the scaling exponent $\nu_\alpha$ of 
Im$\Sigma_\alpha(\mathrm{i}\omega_n)$.
Note that a Fermi liquid exhibits $\nu_\alpha = 1$ 
whereas NFL generally exhibits $\nu_\alpha < 1$.

In general, calculations for finite-size systems would suffer from artificial effects due to the finite level spacing. To mitigate these effects,
we take thermal average of scattering events considering finite but small temperature. 
This requires
the self-energy on the real-frequency axis,
which can be obtained through
the analytic continuation, $\mathrm{i}\omega_n\to\omega + \mathrm{i}\eta$, where $\eta$ is a small broadening. 
We
set $\eta=0.002\approx W_b/N$, where $W_b$ is the bandwidth for $U=0$. In the quasiparticle energy basis, the onsite Hubbard interaction yields a four-point vertex diagram~\cite{Fetter1971}. The self-energy on the real-frequency axis is given by
\begin{align}
\label{realfreqselfgeneral}
\Sigma_{\alpha\beta}(\omega+\mathrm{i}\eta)=\sum_{\lambda,\zeta,\xi}
\frac{
\mathcal{W}_{\alpha\lambda;\zeta\xi}\mathcal{W}_{\beta\delta;\zeta\xi}^*
\,F_{\zeta\xi\lambda}
}{
\omega + \mathrm{i}\eta - \epsilon_\zeta-\epsilon_\xi + \epsilon_\lambda
},
\end{align}
where
\begin{equation}
\mathcal{W}_{\alpha\lambda;\zeta\xi}
=U\sum_i\psi_\alpha(i)^*\psi_\lambda(i)^*\psi_\zeta(i)\psi_\xi(i),
\end{equation}
and
\begin{equation}
F_{\zeta\xi\lambda}
=
(1-f_\zeta)(1-f_\xi)f_\lambda
+
f_\zeta f_\xi(1-f_\lambda).
\end{equation}
We focus
on the diagonal contributions of self-energy matrix in  Eq.~\eqref{realfreqselfgeneral}, i.e., $\Sigma_{\alpha}\equiv\Sigma_{\alpha\alpha}$.
The imaginary part of this self-energy defines the scattering rate of each quasiparticle state, $\psi_{\alpha}$, as
\begin{equation}
\Gamma_\alpha(\omega) = -\mathrm{Im}\,\Sigma_\alpha(\omega + i\eta),
\end{equation}
and the low-energy scaling is extracted from
\begin{equation}
\Gamma_\alpha(\omega) - \Gamma_\alpha(\omega\to0) \sim |\omega|^\nu.
\end{equation}
Here, $\Gamma_\alpha(\omega\to0)=\lim_{\omega\to0}\Gamma_\alpha(\omega)>0$, and $\nu<2$ indicates 
NFL
characteristics.

Since we introduce a finite temperature which broadens the discrete energy levels to mitigate finite-size effects, the electronic state is no longer a pure quasiparticle state, but rather a thermal state composed of a mixture of states within an energy window of the order of the temperature around the Fermi level. Accordingly, the quasiparticle lifetime would be defined as a thermal average over the contributions from individual states. 
Specifically, we
consider thermal averaged 
scattering rate
given by
\begin{align}
\label{gammaT}
\Gamma_T(\omega)&\equiv\mathrm{Im}\Sigma_{T}(\omega,T)\nonumber\\
&=\left[\int_{-\infty}^{+\infty}d\epsilon_\alpha \Gamma_\alpha(\omega)^{-1}\frac{\partial f(\epsilon_\alpha,T)}{\partial \epsilon_\alpha}\right]^{-1}.
\end{align}
We explore the NFL characteristics in terms of the scaling behaviors of $\Gamma_T(|\omega|\ll1)$.

\section{Results}
\label{sec:results}
\subsection{Thermodynamic property}
\begin{figure}[h]
	\centering
	\includegraphics[width=0.5\textwidth]{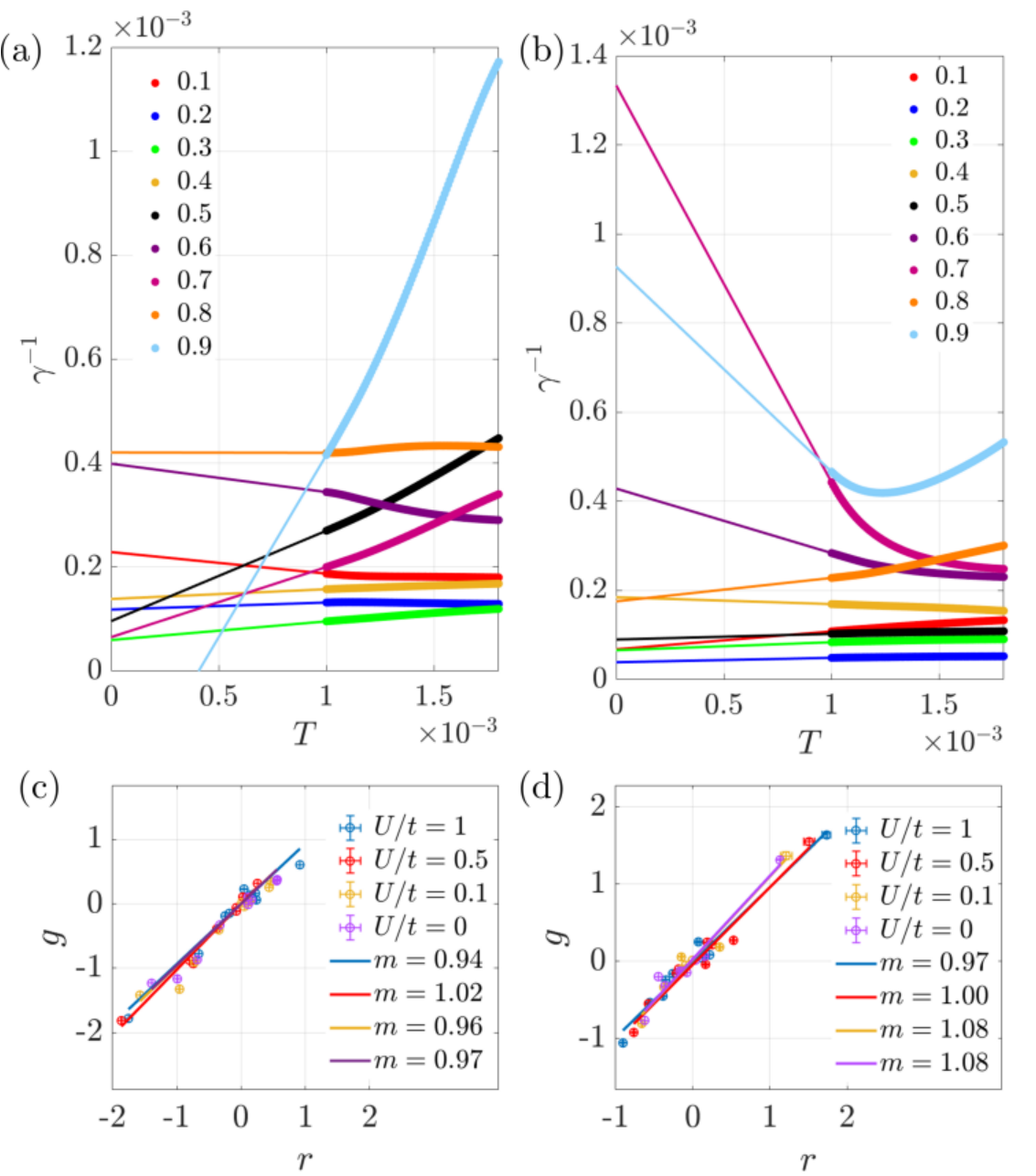}
	\caption{(a,b) Inverse of the effective Sommerfeld coefficient 
    plotted against temperature at $U=0.5t$. Different colors represent different filling fractions. 
    Here, the results 
    are linearly extrapolated to $T<10^{-3}$. (c,d) Relationship between the exponent $g$ and $r$ characterizing the temperature dependence of $\gamma$ and $R$, respectively.
    Different colored lines represent $g=mr$ relationship for different values of $U/t$. Panels (a,c) [(b,d)] display the results for the Penrose (Ammann--Beenker) tiling.
    }
	\label{fig:gamma}
	
\end{figure}

We first examine the 
temperature dependence
of the specific heat. Figures~\ref{fig:gamma}(a,b) show the effective Sommerfeld coefficient, \(\gamma=C(T)/T\), for the Penrose and Ammann--Beenker tilings, respectively, as a function of temperature at various filling fractions. For generic filling fractions, \(\gamma\) exhibits a pronounced temperature dependence, in clear contrast to the constant behavior 
of the Fermi liquid: 
At low temperatures, \(\gamma\) can increase or decrease as $T$ decreases, depending on the filling fraction.
Remarkably, for several filling fractions, \(\gamma\) increases anomalously upon lowering the temperature, and at particular filling fractions it even exhibits divergent critical behavior in the low temperature limit [see 90$\%$ filling fraction of Penrose tiling shown in Fig.~\ref{fig:gamma}(a), for instance]. This demonstrates that the observed NFL behavior of \(\gamma\) is not merely a conventional insulating tendency associated with an increasing \(\gamma(T)\), but rather reflects an anomalous critical property. 
Namely, this
unconventional temperature dependence originates from the nontrivial distribution of energy eigenvalues and the fragmented DOS associated with the critical states inherent to quasicrystalline structures. These results indicate that, even in the weak-interaction regime, quasicrystals can give rise to 
diverse anomalous
thermodynamic behaviors and, furthermore, can induce a low-temperature divergent criticality of \(\gamma\) at some filling fractions.

To demonstrate
that the fragmented spectrum 
gives rise to
the anomalous temperature dependence of 
$\gamma(T)$,
we consider
the \(R\)-function,
\begin{align}
\label{R-function}
R=\rho+T\frac{d\rho}{dT},
\end{align}
associated with the thermal DOS, $\rho=\rho(T)$ in Eq.~\eqref{thermaldos}.
Roughly speaking, the low-temperature behavior of the entropy, \(S \sim T\rho\), will connect $\gamma(T)$ with $R$. 
To make this statement more precise, we take into account that 
in quasicrystals, the 
DOS at the Fermi level can exhibit anomalous temperature dependence beyond the simple thermal broadening of the Fermi--Dirac distribution function. 
This originates from the combined effects of the intrinsically fragmented spectrum of the quasicrystal and the inhomogeneous temperature evolution of the self-consistent charge density, which consequently modifies the Hartree-shifted mean-field Hamiltonian in Eq.~\eqref{HF}. Note that the local charge density and the associated Hartree-shift inhomogeneity exhibit strong filling dependence, while their temperature evolution develops a fractal spatial pattern inherited from the fractal LDOS map (see Appendix~\ref{A_add}).

Figures~\ref{fig:gamma}(c,d) 
plot
the exponent $g$ defined in Eq.~\eqref{g_exponent}
against the exponent 
\begin{align}
\label{r_exponent}
r=\frac{\partial\log R}{\partial \log T}
\end{align}
(i.e., $R\propto T^r$).
We find
a positive correlation between them for generic filling fractions in both tilings. Specificially, in the weak interaction regime, we obtain \(g \sim mr\) with \(m\) close to unity. Thus, we have $\gamma\propto R$ relation. This demonstrates that, even in quasicrystalline systems, the entropy and specific heat are fundamentally governed by the 
DOS
near the Fermi level, while the fragmented spectrum characteristic of quasicrystals leads to anomalous temperature dependence of the 
DOS
and specific heat. 
Hence, by examining the temperature dependence of the \(R\)-function associated with the 
DOS
near the Fermi level, one can predict the NFL thermodynamic behaviors through the relation $g\sim mr$ with $m\approx 1$, implying $\gamma\propto R$ relation.

\subsection{Magnetic response}

\begin{figure}[h]
	\centering
	\includegraphics[width=0.5\textwidth]{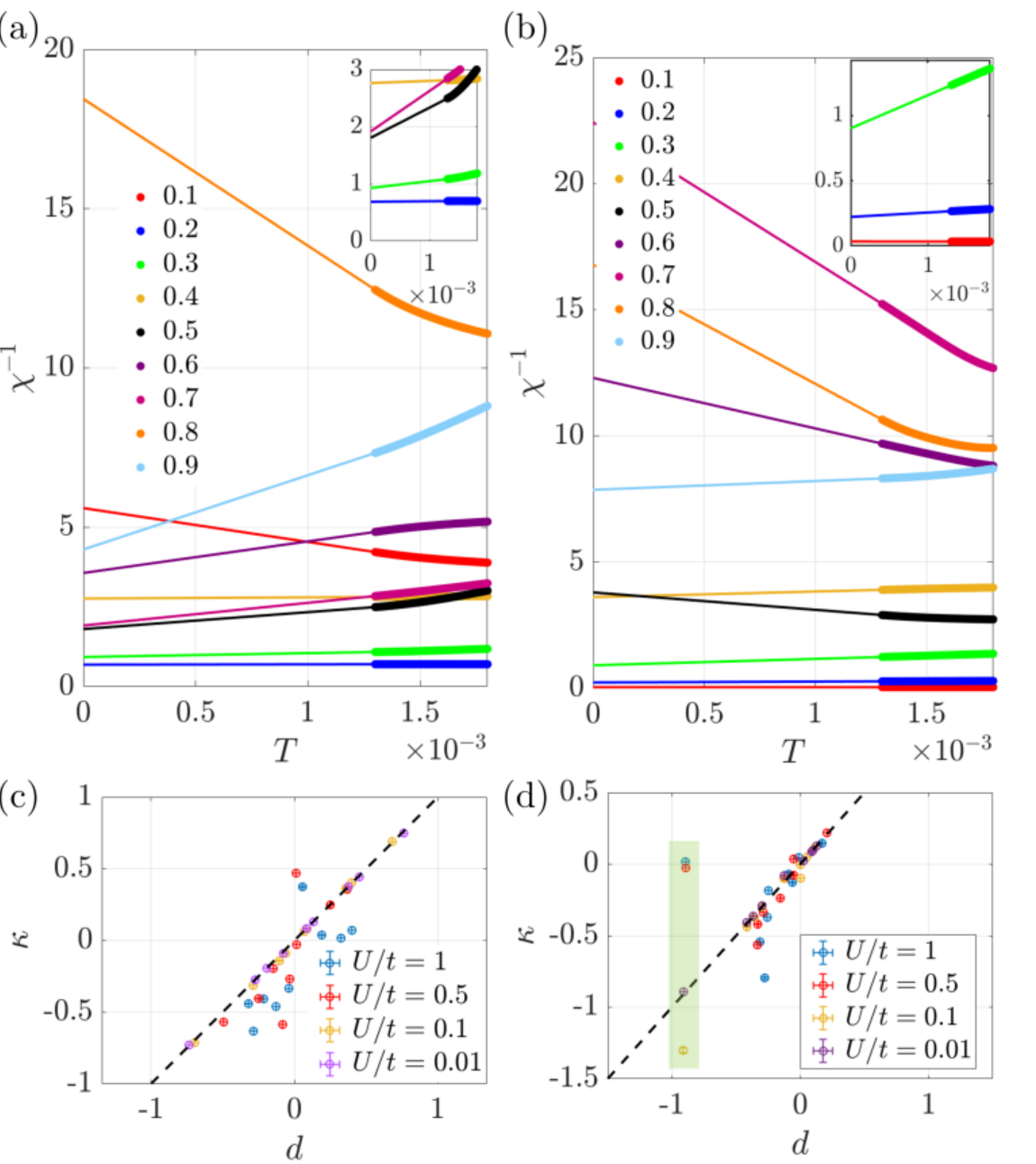}
	\caption{(a,b) Inverse magnetic susceptibility plotted against 
    temperature. Different colors represent different filling fractions. Panel (a) shows the results for the Penrose tiling, while panel (b) corresponds to the Ammann--Beenker tiling. Here, $U=0.5t$. The insets are drawn to emphasize the behaviors for large susceptibilities. Here, the results for $T<10^{-3}$ is obtained by using linear extrapolation. (c,d) Relationship between the exponents $\kappa$ and $d$ characterizing the temperature dependence of 
    $\chi$ and $\rho$,
    respectively. Panel (c)[(d)] displays the result for the Penrose tiling
    (the Ammann--Beenker tiling). Black dashed lines 
    represent
    $\kappa=d$ relationship. Green shaded region in (d) is drawn to emphasize large deviations from the $\kappa=d$ relationship at $10\%$ filling even for small $U$ values. 
    }
	\label{fig:suscep}
	
\end{figure}

Next, let us investigate 
the magnetic response in the quasicrystals.
Figures~\ref{fig:suscep}(a,b) show the magnetic susceptibility, \(\chi\), for the Penrose and Ammann--Beenker tilings, respectively, as a function of temperature at various filling fractions. Similarly to the behavior observed in the specific heat, \(\chi\) exhibits pronounced temperature dependence over a wide range of filling fractions, deviating from the nearly temperature\mbox{-}independent Pauli susceptibility expected in 
the Fermi liquids. We note that the paramagnetic response 
in quasicrystals is spatially inhomogeneous 
due to
the inhomogeneous 
LDOS
(see Appendix~\ref{A3}). In detail, the thermal 
LDOS,
$\rho_i(T)$, and the local magnetization induced by a uniform external field show a positive correlation. This suggests that the magnetic susceptibility can also display anomalous 
temperature dependence. 

Similarly to the specific heat, 
this anomalous
temperature dependence of the magnetic susceptibility can be understood from the fragmented 
DOS
associated with the critical states of quasicrystals. Moreover, owing to the self\mbox{-}consistent Hartree shift in Eq.~\eqref{HF}, the effective mean\mbox{-}field Hamiltonian itself acquires additional temperature dependence through the evolution of the charge-density distribution, enhancing the unconventional magnetic response beyond simple thermal broadening effects.

Figures~\ref{fig:suscep}(c,d) show the relation between the exponent $\kappa$ [Eq.~\eqref{kappa_exponent}] characterizing the temperature dependence of the magnetic susceptibility and the exponent $d=\frac{\partial \log \rho}{\partial \log T}$ associated with the temperature dependence of the 
DOS,
$\rho$.
Similarly to the specific heat coefficient, we find a linear relation between the two exponents, i.e., $\kappa\approx d$, for almost all filling fractions in both tilings. In the weak interaction regime, 
$\kappa$
is approximately proportional to 
$d$
with a proportionality coefficient close to unity for almost all 
the filling fractions. Note that $\kappa\sim d$ indicates that $\chi\propto\rho$ relation. This correspondence demonstrates that the anomalous magnetic response is fundamentally governed by the singular 
DOS
near the Fermi level induced by quasicrystalline criticality. Therefore, the temperature dependence of the 
DOS
provides a useful indicator not only for 
NFL
thermodynamic behavior in the specific heat but also for anomalous magnetic responses in interacting quasicrystals.

Before proceeding further, we note that even in the weak\mbox{-}interaction regime, there exist particular filling fractions that strongly deviate from the universal linear relation between \(\kappa\) and \(d\) discussed above. An 
example is the \(10\%\)-filled Ammann--Beenker tiling [the data points inside of green shaded area in Fig.~\ref{fig:suscep}(d)]. It turns out that this deviation originates from a singular DOS induced by confined localized states near the Fermi level~\cite{Arai1988,Jeon2022,Mirzhalilov2020}. Note that near singular structures in the 
DOS,
such as van Hove singularities or flat bands, interaction effects can be significantly enhanced even for weak interactions due to the large 
DOS.
In particular, according to the Stoner criterion~\cite{Stoner1938}, systems with such a singular 
DOS
can develop magnetic ordering even at small \(U\), significantly changing the behavior of the magnetic susceptability~\cite{Koga2017,Koga2020}. We verify this scenario by exploring 
a magnetically ordered
solution for the \(10\%\)-filled Ammann--Beenker tiling that largely deviates from the linear scaling relation even for small $U/t=0.1$ (see Appendix~\ref{A4}).

Note that 
for $U/t\gtrsim 1$,
magnetic orders are favored at some filling fractions
even in the absence of singularities. At these filling fractions, the 
tendency to
magnetic ordering 
leads to deviations from the linear relation between \(\kappa\) and \(d\). However, unlike the case driven by confined localized states~\cite{Arai1988}, such magnetic ordering disappears 
for a small $U$,
where the universal linear relation between \(\kappa\) and \(d\) is recovered. 
Thus,
depending on the interaction strength, most filling fractions in quasicrystals can host either paramagnetic (weak $U$) or magnetically ordered states (intermediate or strong $U$) (see Appendix~\ref{A4}).

\subsection{Wilson ratio}

\begin{figure}[h]
	\centering
	\includegraphics[width=0.5\textwidth]{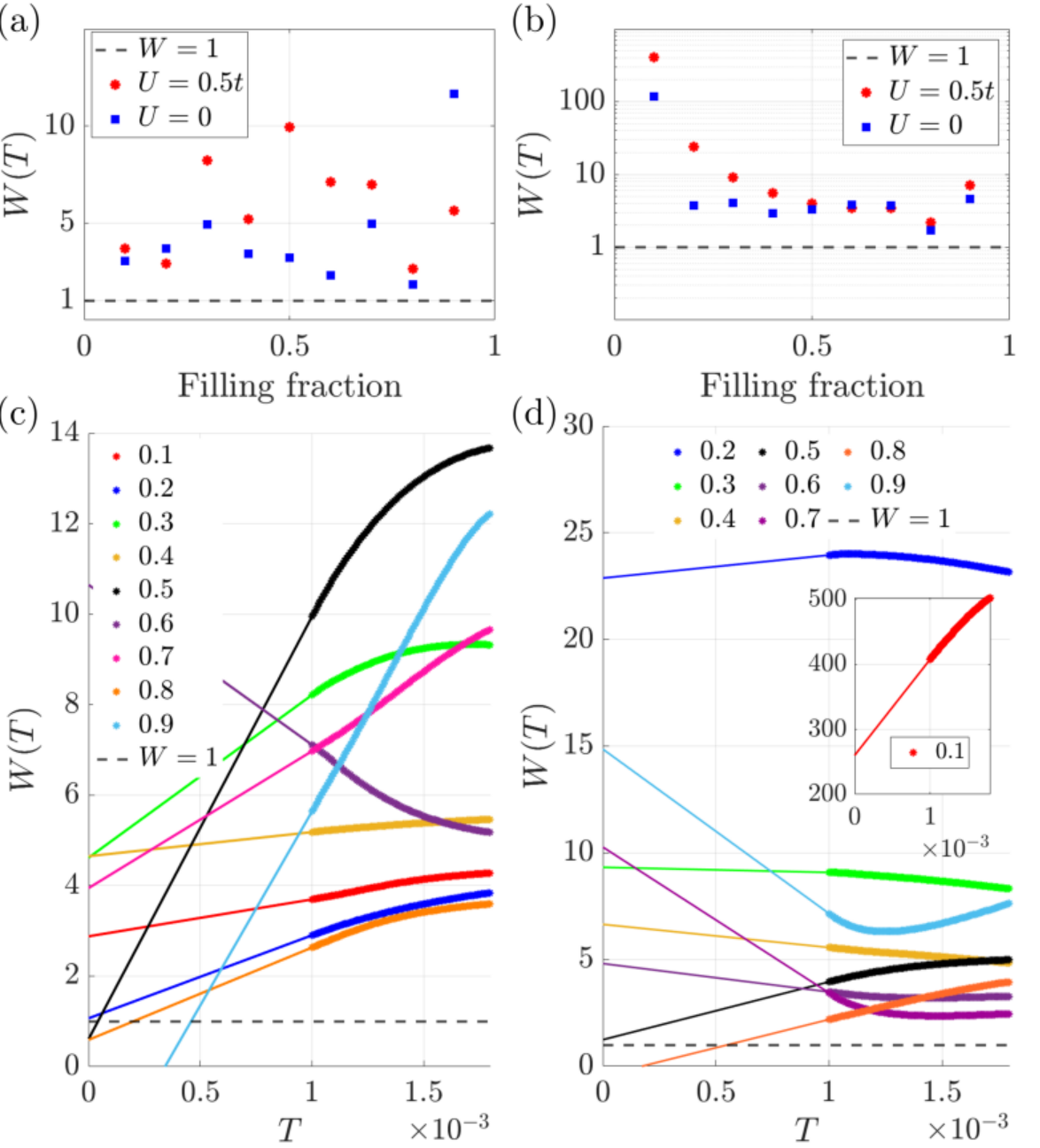}
	\caption{Wilson ratio at $T=10^{-3}$ for different filling fractions of (a) the Penrose and (b) the Ammann--Beenker tilings. Wilson ratio for $U=0.5t$ as a function of temperature for different filling fractions of (c) the Penrose tiling and (d) the Ammann--Beenker tiling. Here, the results for $T<10^{-3}$ are obtained by using a linear extrapolation. 
    Inset to (d) shows the result for 10\% filling.}
	\label{fig:Wilson}
	
\end{figure}

Now, let us investigate the Wilson ratio, $W(T)$ in Eq.~\eqref{Wilsonratio} to gain further insight into the low-energy electronic state. Figures~\ref{fig:Wilson}(a,b) show the Wilson ratio for various filling fractions at a low temperature, $T=10^{-3}t$. 
We find $W(T)>1$ for all the parameters we studied, even for $U=0$.
Notably, the magnitude of the Wilson ratio varies substantially with filling fraction, reflecting the corresponding changes in the 
DOS
near the Fermi level. For instance, the confined localized states at $10\%$-filled Ammann-Beenker tiling lead to anomalously large Wilson ratio of more than 100 [see Fig.~\ref{fig:Wilson}(b)].

For $U>0$, the Wilson ratio 
increases, compared to the $U=0$ value, at most filling fractions since the interactions enhance the magneitc fluctuations, while decreases for some filling fractions. The filling-dependent effect of $U$ originates from the changes in the electronic spectrum and states induced by the inhomogeneous Hartree shift. Hence, the intrinsically large Wilson ratio primarily originates from the quasiperiodic electronic structure and the associated 
DOS, whereas electron interactions may 
further change
the strength of magnetic fluctuations 
through inhomogeneous Hartree shift.

Figures~\ref{fig:Wilson}(c,d) 
plot
the Wilson ratio for $U=0.5t$ and different filling fractions as a function of temperature. 
Unlike the Fermi liquid, the calculated 
Wilson ratio 
exhibits a significant temperature dependence.
This will be attributed to
the temperature dependence of the 
thermal DOS, $\rho(T)$. 
Interestingly,
the temperature dependence of the Wilson ratio exhibits diverse trends:
It increases or decreases upon cooling, depending on the underlying tiling pattern and filling fraction. Notably, the zero-temperature Wilson ratio, estimated by a linear extrapolation, exceeds unity for most filling fractions, even in some cases where it decreases upon cooling. This indicates 
enhanced 
spin fluctuations 
across most filling fractions. On the other hand, for some filling fractions such as the $90\%$-filled Penrose tiling and the $80\%$-filled Ammann--Beenker tiling, the extrapolation suggests that the Wilson ratio would be strongly suppressed upon cooling and 
be smaller than one
at low temperatures [see skyblue and orange curves in Figs.~\ref{fig:Wilson}(c,d), respectively]. This extrapolation implies that the low-temperature entropy may not be accompanied by a comparable enhancement of the magnetic response at these particular filling fractions, and raises the possibility that weakly magnetic low-energy excitations could become dominant upon further cooling.

\subsection{Transport property}
\subsubsection{Scaling of scattering rate and residual resistivity}
Having discussed the anomalous behaviors observed in 
thermodynamic and magnetic properties,
we now turn to the transport signatures manifested in the anomalous self-energy scaling and residual resistivity~\cite{kimura1989electronic,PhysRevLett.70.3915,roche1998fermi}. Remind that within the conventional Landau Fermi-liquid framework, the imaginary part of the self-energy is 
supposed
to follow the quadratic scaling relation, $\mathrm{Im}\,\Sigma(\omega,T)\propto \omega^2+\pi^2T^2$,
which reflects the existence of long-lived quasiparticle excitations in the low-energy limit~\cite{baym2008landau}. Correspondingly, the electrical resistivity exhibits the characteristic \(T^2\)-dependence arising from electron--electron scattering processes. Deviations from these canonical scaling forms therefore provide direct evidence for the breakdown of coherent quasiparticle behavior and emergence of NFL state~\cite{deguchi2012quantum,Sato2022effects}. Here, we 
focus on
the Penrose tiling as a concrete platform to explore NFL transport characteristics. However, 
similar observations could be found in the Ammann--Beenker tiling (see Appendix~\ref{A5}).

\begin{figure}[h]
	\centering
	\includegraphics[width=0.5\textwidth]{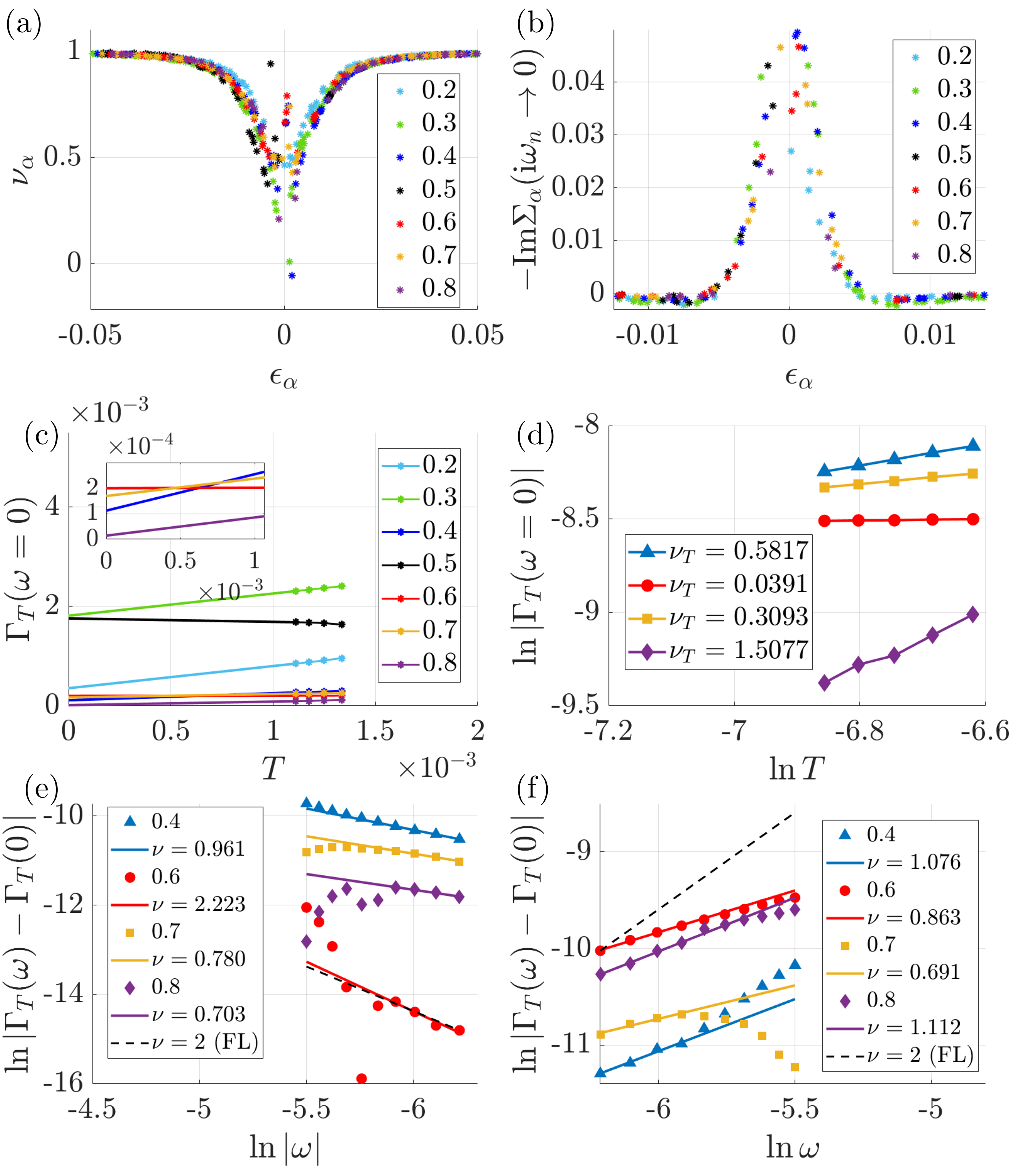}
	\caption{NFL behaviors of thermal scattering rate of Penrose tiling for $U/t=1$. (a) Scaling exponent, $\nu_\alpha$ of imaginary part of 
    self-energy as a function of Matsubara frequency (b) residual scattering rate for the quasiparticle state, $\psi_{\alpha}$ as a function of energy, $\epsilon_\alpha$. $\nu_\alpha=1$ and $-\mathrm{Im}\Sigma_{\alpha}(\mathrm{i}\omega_n\to0)=0$ correspond to Fermi liquid, while $\nu_\alpha<1$ and $-\mathrm{Im}\Sigma_{\alpha}(\mathrm{i}\omega_n\to0)>0$ indicate NFL behavior. Different colors represent different filling fractions. Calculations were performed at $T=10^{-3}t$. (c) Scattering rate at $\omega=0$, $\Gamma_T(\omega=0)$, as a function of temperature for different filling fractions. The inset is drawn to emphasize small values of $\lim_{T\to0}\Gamma_T(0)$. Except $80\%$ filling (violet), general filling fractions clearly admit nonzero $\lim_{T\to0}\Gamma_T(0)$. (d) Log-log plot of zero-frequency scattering rate as a function of temperature. Blue, red, gold and violet colors represent different filling fractions, $0.4,0.6,0.7$ and $0.8$, respectively,
    where a
    paramagnetic solution is stabilized.  (e,f) Scaling behaviors of thermal scattering rate at $T=10^{-3}t$ as a function of frequency. Different colors represent different filling fractions. Black dashed line 
    represents
    Fermi liquid scaling exponent, $\nu=2$. Panels (e) and (f) show negative and positive frequency results, respectively. 
    }
	\label{fig:scatteringscaling}
	
\end{figure}

We first consider the scattering rate of each quasiparticle state, $\psi_\alpha$ of the Hartree-shifted Hamiltonian in Eq.~\eqref{HF} on the Penrose tiling. By using the self-energy as a function of Matsubara frequency, Eq.~\eqref{energybasisselfMatsubara}, we investigate its scaling exponent, $\nu_\alpha$, and residual scattering rate, $\mathrm{Im}\Sigma_{\alpha}(\mathrm{i}\omega_n\to0)$. Figures~\ref{fig:scatteringscaling}(a,b) illustrate $\nu_\alpha$ and $\mathrm{Im}\Sigma_{\alpha}(\mathrm{i}\omega_n\to0)$ as functions of energy, $\epsilon_\alpha$, for general filling fractions.
Notably, 
$\nu_\alpha$ deviates from 1 and $\mathrm{Im}\Sigma_{\alpha}(\mathrm{i}\omega_n\to0)$ becomes nonzero
predominantly around the Fermi energy, $\epsilon_\alpha=0$, exhibiting NFL behaviors:
The Fermi-liquid values of $\nu_\alpha=1$ and $\mathrm{Im}\Sigma_{\alpha}(\mathrm{i}\omega_n\to0)=0$ are recovered 
only at energies far away from the Fermi energy.
This is because 
the deviation from $\nu_\alpha=1$
originates from singular low-energy scattering processes and therefore can emerge only for states sufficiently close to the Fermi energy.

Next, let us consider a thermal scattering rate, $\Gamma_T(\omega)$, given by Eq.~\eqref{gammaT} to investigate NFL characteristics encoded in self-energy at finite temperatures. Specifically, a general thermal scattering rate of NFL would have the following scaling forms for small $\omega$ and low $T$,
\begin{align}
\label{generalgammanonFL}
\Gamma_T(\omega)&=\Gamma_T(0)+\vert\omega\vert^\nu, \\
\nu_T&\equiv \frac{\partial \log \Gamma_T(0)}{\partial \log T}
\end{align}
with $\nu,\nu_T<2$. In particular, quasicrystalline NFL states would have residual scattering rate as $\lim_{T\to 0}\Gamma_T(0)>0$ without thermal or energy excitations. It is worth noting that quasicrystalline systems are experimentally known to exhibit finite residual resistivity, 
which increases as the sample quality increases \cite{Rapp2008}, suggesting that it originates
from their intrinsically aperiodic lattice structures~\cite{,kimura1989electronic}.
The absence of translational symmetry can induce unconventional electronic scattering even in the absence of extrinsic disorder, leading to suppressed charge transport at low temperatures even compared to the case of 
disorder~\cite{kimura1989electronic,roche1997electronic}. Such behavior has been experimentally observed in transport measurements on various quasicrystalline compounds~\cite{kimura1989electronic,PhysRevB.59.308,Dolinsek2012}.


Figure~\ref{fig:scatteringscaling}(c) exhibits the thermal scattering rate for $\omega=0$ as a function of temperature for various filling fractions at $U=t$. Notably, most of filling fractions admit finite scattering rate for low temperatures as $\lim_{T\to 0}\Gamma_T(0)>0$ shows. This implies that quasicrystalline structures would 
produce
residual resistivities 
through electron-electron interactions
for generic filling fractions as the finite residual lifetimes of quasiparticle states.

Focusing on the paramagnetic states, we consider four different filling fractions (0.4, 0.6, 0.7 and 0.8) on the Penrose tiling at $U/t=1$. Figure~\ref{fig:scatteringscaling}(d) emphasizes anomalously small scaling exponents, $\nu_T<2$ for these filling fractions. Thus, these states admit larger thermal 
scattering rate
compared to 
Fermi-liquid states at low temperature. 

Figures~\ref{fig:scatteringscaling}(e,f) exhibit the thermal scattering rate at a low temperature as a function of negative and positive frequencies, respectively. First, the scaling exponent $\nu$ of $\Gamma_T(\omega)$ 
is generally smaller than 2 in the low-frequency regime. 
This indicates that the resistivity, which is related to the scattering rate, would show anomalously small scaling exponents. Note that the scaling exponent varies with filling and can exhibit diverse NFL values. Second, the frequency dependence of $\Gamma_T(\omega)$ is 
strongly asymmetric w.r.t. $\omega$ for general filling fractions. This reflects the fragmented 
and non-smooth 
DOS
of quasicrystals. Lastly, these features reveal an even more intriguing consequence: NFL self-energy scaling in quasicrystals generally exhibits distinct scaling exponents 
w.r.t. $\omega$ and $T$,
i.e., $\nu\neq\nu_T$. Such distinct responses can be experimentally accessed through photoemission spectroscopy~\cite{Hufner2003,Nozue2024,Sakamoto2026} and DC conductivity measurements~\cite{Rotenberg2008,Ziman1960,Stewart2001,roche1997electronic}, respectively.

Before going further, we note that 2D Fermi liquid has some $\log$-correction of self-energy, $\mathrm{Im}\Sigma(\omega)\approx \omega^2\log\vert\omega\vert$ because low-energy quasiparticle scattering near the Fermi surface has a marginally singular phase space. This alters the scaling exponent of Fermi liquid as $\nu=2$ to $2+(\log \vert\omega\vert)^{-1}$~\cite{Hodges1971,Bloom1975,Giuliani2005,Chubukov2003}. Nevertheless, we emphasize the conclusion that the 
self-energy shows a NFL behavior
as inferred from Figs.~\ref{fig:scatteringscaling}(d–f), remains unchanged even we take into account the logarithmic corrections characteristic of 2D systems. 

\subsubsection{Real-space analysis}
Let us turn to a real-space analysis of the 
scattering rate shown
in Fig.~\ref{fig:scatteringscaling}. While the scaling analysis of the scattering rate captures the unconventional metallic response at a phenomenological level, understanding its microscopic origin in quasicrystals is crucial to 
obtain
a complementary viewpoint beyond momentum-space descriptions. Owing to the absence of translational symmetry, electronic states in quasicrystals are inherently sensitive to local geometrical structures and spatial inhomogeneity, making real-space analysis suited for investigating the emergence of NFL behavior in these systems.

\begin{figure}[h]
	\centering
	\includegraphics[width=0.5\textwidth]{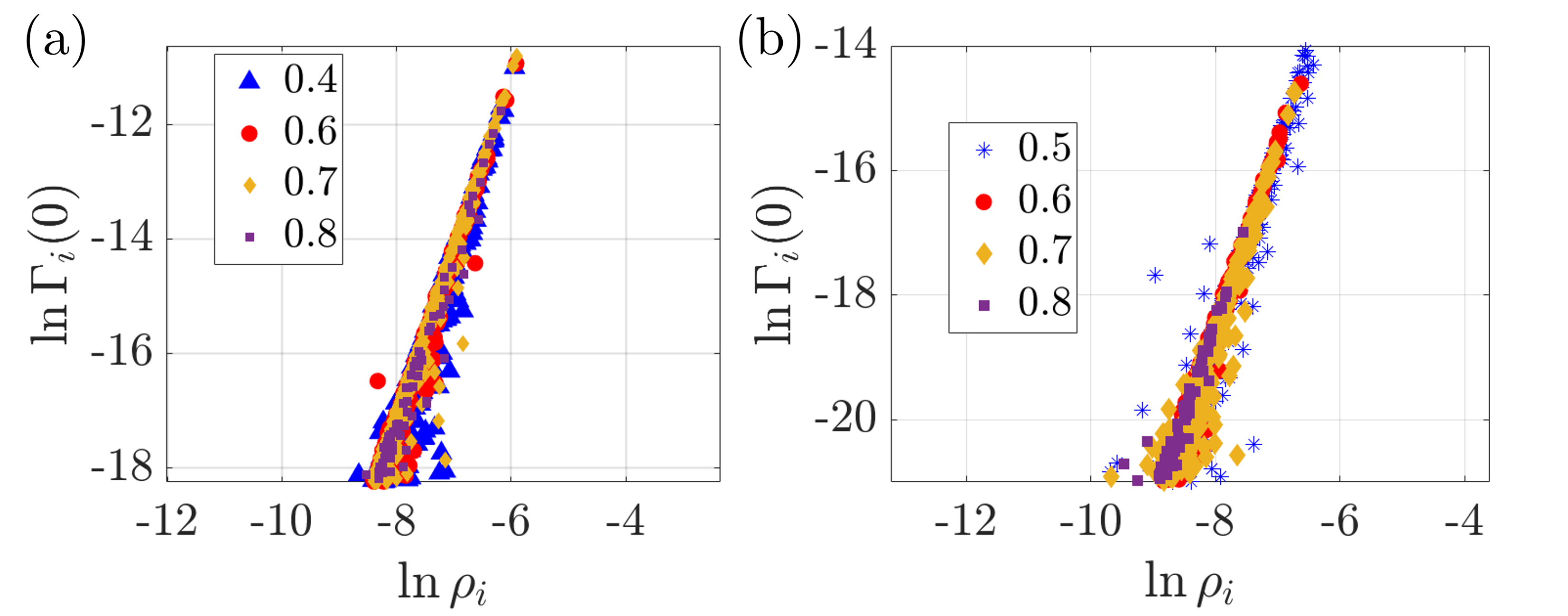}
	\caption{
    Relationship between 
    LDOS
    $\rho_i$ at the Fermi level and local residual resistivity $\Gamma_i(0)$, calculated at $T=10^{-3}t$ for different filling fractions in (a) the Penrose tiling with $U/t=1$ and (b) the Ammann--Beenker tiling with $U/t=0.5$. $\Gamma_i(0)\approx\rho_i^3$ is found for general filling fractions 
    on both tilings.
    }
	\label{fig:realspace}
	
\end{figure}

We note that a large 
LDOS
drives the local self-energy scaling exponent and residual scattering rate further away from the Fermi-liquid values. This is because the NFL behavior stems from the states around the Fermi energy. 
In Fig.~\ref{fig:realspace}, we find that
the local residual scattering rate and 
LDOS
follow a general relationship, $\Gamma_i(0)\approx \rho_i^3$, for various filling fractions and tiling patterns.
This relationship can be traced back to the second-order self-energy, which is given by the convolution of three Green's functions [see Eq.~\eqref{perturbativeselflocal}].

\begin{figure*}[!t]
	\centering
	\includegraphics[width=1.0\textwidth]{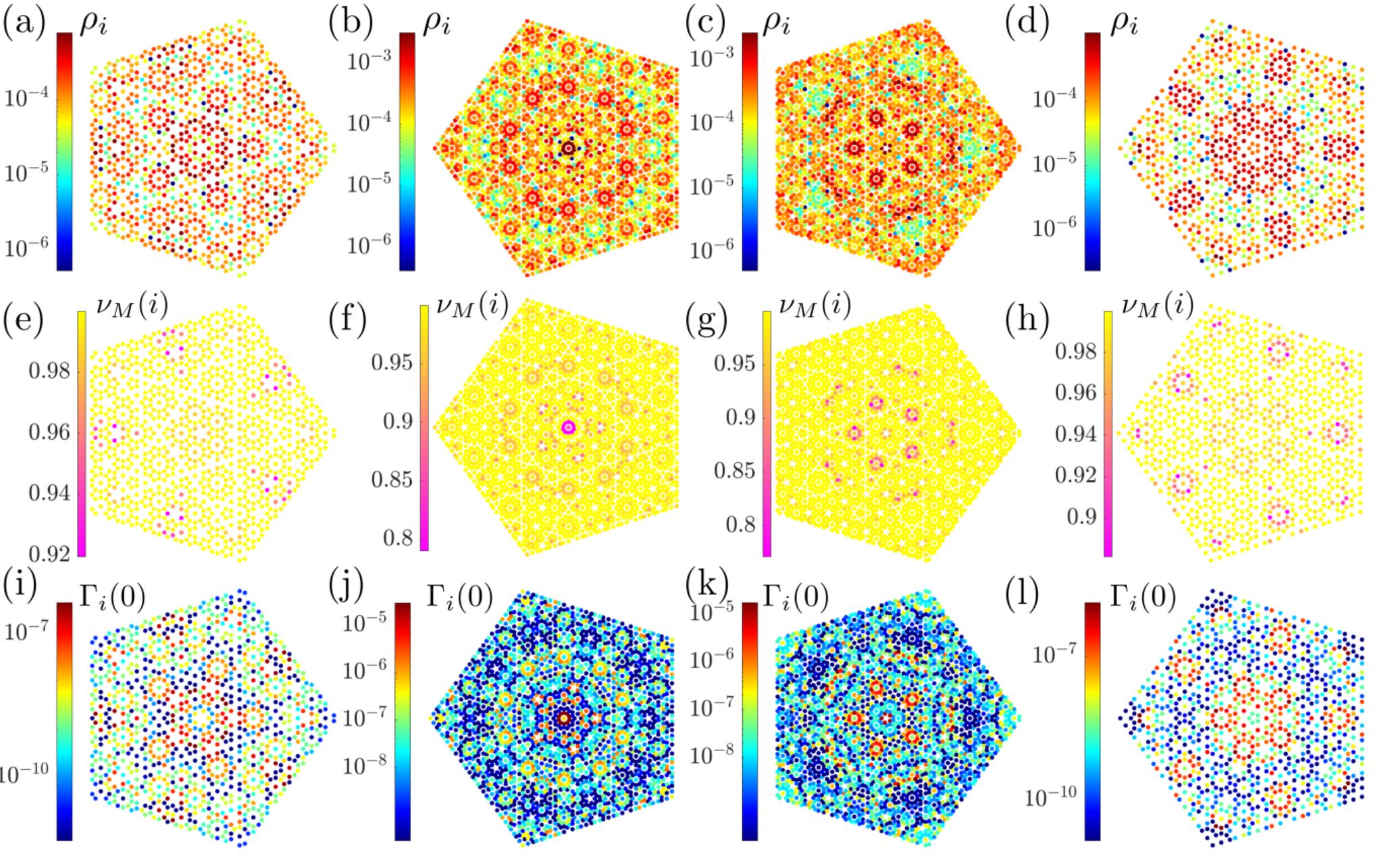}
	\caption{Perpendicular-space profile of (a-d) LDOS, $\rho_i$, at the Fermi level, (e-h) scaling exponent of local Matsubara-frequency self-energy, $\nu_M(i)$, and (i-l) local residual scattering rate, $\Gamma_i(0)$, of the Penrose tiling for $U=t$, 60\% filling, and $T=10^{-3}t$. 
    }
	\label{fig:perpspace}
\end{figure*}

Since the electronic states in quasicrystals are generally critical~\cite{Tokihiro1988}, the wave functions and 
LDOS
near the Fermi energy that drive the NFL behavior are expected to be highly inhomogeneous. In particular, these critical states show long-range correlations.
This suggests that local NFL indicators, such as $\nu_M(i)$ and $\Gamma_i(0)$, are highly inhomogeneous and cannot be explained solely by simple local geometrical features such as connectivity. To show this, we 
plot
$\nu_M(i)$, $\Gamma_i(0)$, and the 
LDOS
in perpendicular space. Figure~\ref{fig:perpspace} displays that all of these quantities 
feature
long-range correlations, as they show fine structures in the perpendicular space.
Note that both $\nu_M(i)$ and $\Gamma_i(0)$ 
show particularly large deviations 
from the Fermi-liquid values at the sites with large 
$\rho_i$.

\section{Discussion}
\label{sec:discussion}
We have studied the characteristics of the weak-coupling NFL state
in 2D quasicrystals.
Strong electron correlations are not a necessary prerequisite for anomalous metallic behavior in
these
systems. Instead, the critical single-particle wavefunctions and fragmented spectra intrinsic to quasicrystals already provide a singular background to the NFL behavior.
We have revealed that, 
on 
these electronic structure,
even weak interactions generate 
further anomalous behaviors in 
thermodynamic, magnetic, and transport properties. 
This distinguishes the NFLs
in quasicrystals from more conventional ones
based on
magnetic quantum criticality~\cite{vonLohneysen2007,gegenwart2008quantum}, Kondo breakdown~\cite{Si2001}, or strong-coupling collective fluctuations~\cite{Varma1989,Moriya1985}.

A key outcome of our result is that the 
NFL
behavior of quascirystals is controlled by the DOS, and in particular by the LDOS near the Fermi energy. The anomalous temperature dependence of thermodynamic quantities arises from the unconventional low-energy DOS together with its filling-dependent reconstruction induced by the inhomogeneous temperature dependence of the Hartree shift, while the residual scattering rate and local magnetizations are strongly amplified at sites with enhanced LDOS. 
These behaviors originate from the multifractal critical wavefunctions, which convert
quasiperiodic criticality directly into spatially inhomogeneous electronic responses. Therefore, the NFL behavior in quasicrystals can be broadly controlled by the filling fraction through its influence on the critical states near the Fermi energy. This offers a unique mechanism for engineering anomalous electronic properties, with distinct signatures appearing in thermodynamic and transport quantities such as the specific heat, magnetic susceptibility, and electrical resistivity.

The Wilson ratio provides a diagnostic of the nature of low-energy excitations. For most filling fractions, 
we have obtained values 
larger than unity at low temperatures. 
The onsite interaction $U>0$ further increases the value for most filling fractions, 
indicating that these NFL behaviors are 
accompanied by an enhanced magnetic response. 
Remarkably,
the calculated Wilson ratios are comparable to those measured in heavy-fermion quasicrystals~\cite{deguchi2012quantum,Watanuki2014}. Since our results are obtained for weakly-correlated regime, this suggests that the NFL behavior observed in the experiment may be rooted in the intrinsic quasiperiodic electronic structure rather than strong electron correlations~\cite{Andrade2015}. We further find that both the magnitude of the Wilson ratio and its temperature dependence vary substantially with filling fraction, exhibiting qualitatively distinct behaviors such as increasing or decreasing upon cooling. This diversity indicates that the NFL behavior of quasicrystal is highly tunable through changes in the filling fraction, reflecting the corresponding evolution of the DOS near the Fermi level.

Our analysis of the transport property 
is based 
on the diagonal component of self-energy. We emphasize that the anomalous scaling exponents and finite residual scattering rate emerging 
from this diagonal self-energy is significant because it 
represents the contribution of the critical states and fragmented spectrum 
at the simplest nontrivial perturbative level.
The anomalous self-energy scaling obtained here provides direct evidence for 
transport behavior. However, the DC conductivity is not determined solely by the self-energy, since current matrix elements and vertex corrections can modify the relation between the single-particle and transport lifetimes~\cite{Mahan2000,Ashcroft1976,Bruus2004,Prange1964}. A quantitative analysis of conductivity therefore remains an important direction for future work. Nevertheless, the present results already provide a theoretical explanation for experimentally observed transport anomalies in quasicrystals, including finite residual resistivity and anomalous scaling behavior of scattering rate. 

The off-diagonal components of self-energy, $\Sigma_{\alpha\beta}$ with $\alpha\neq\beta$, would be significant in the higher-order corrections with further renormalization of quasiparticle states.
Note that the off-diagonal self-energy components may encode interference between different critical wavefunctions 
and could modify quasiparticle scattering, transport anisotropy, and the spatial propagation of correlations.
Such terms may be particularly important near singular filling fractions, near confined localized states, or in regimes where magnetic or charge instabilities become competitive~\cite{Gonzalez2000,Koga2017,Koga2020}. Hence, including off-diagonal self-energy contributions is also an important direction for future work, though it is currently unfeasible in terms of the computational cost. 

Another important direction for future work is to investigate how the NFL behavior driven by critical states and fragmented spectra in the weak-coupling regime evolves as the interaction strength is increased toward the intermediate- and strong-coupling regimes. While the present study 
has revealed the effects of quasiperiodic criticality on 
thermodynamic, magnetic, and transport properties in the weak-coupling regime,
stronger interactions may qualitatively reshape the low-energy electronic states through enhanced magnetic correlations, local-moment formation, or collective many-body effects~\cite{Stewart2001,deguchi2012quantum,Andrade2015}. Understanding whether the NFL properties established within the weak-coupling regime continuously evolves into heavy-fermion-like~\cite{deguchi2012quantum,Andrade2015}, magnetically ordered~\cite{Koga2017,Koga2020}, or fractionalized phases~\cite{Balents2010}, or whether it is separated from them by interaction-driven phase transitions, remains an open question. Exploring this crossover or transition by using nonperturbative methods, such as real-space dynamical mean-field theory~\cite{georges1996dynamical,Takemori2015,sakai2022doped} and quantum Monte Carlo simulations~\cite{Gubernatis2016} would provide a unified picture of how quasiperiodic criticality and electronic correlations cooperate to produce unconventional quantum states in quasicrystals.

Our
results provide 
several
experimentally testable predictions. First, the temperature dependences of the specific heat and magnetic susceptibility should be tunable 
by shifting the Fermi level through chemical substitution, gating, or pressure. Second, the Wilson ratio should show strong filling and temperature dependences. Third, quasicrystalline NFL states are predicted to exhibit different scaling exponents in their energy ($\omega$) and temperature ($T$) responses. Such a separation is absent in 
Fermi liquids and would be examined using photoemission spectroscopy and DC transport measurements, for instance~\cite{Hufner2003,Nozue2024,Sakamoto2026,Mahan2000,Ashcroft1976,Rotenberg2008,Ziman1960}. Lastly, the residual resistivity should correlate with local spectroscopic signatures of enhanced LDOS. 
This
observation of general finite residual resistivity supports the well-known experimental observation of electron transport in quasicrystals~\cite{roche1997electronic,stadnik1998physical,kimura1989electronic}. Spatially resolved probes, such as (spin-polarized) scanning tunneling spectroscopy, could directly test the predicted correlation between LDOS, anomalous self-energy scaling, and residual scattering as well as local magnetic structures~\cite{Tersoff1985,Bode2003,Wulfhekel2007}. Observing these correlations would provide strong experimental evidence that multifractal critical states are the microscopic origin of NFL behavior in quasicrystals.

A promising platform for testing the present weak-coupling mechanism would be provided by ultracold atoms in optical quasicrystal lattices~\cite{Gross2017,Viebahn2019,Corcovilos2019}. In such systems, the interaction strength can be continuously controlled through the scattering length using Feshbach resonances~\cite{Chin2010}, allowing direct access to the weakly interacting regime considered in this work. Recent experiments have successfully realized 2D quasicrystalline optical lattices with eightfold rotational symmetry closely related to the Ammann--Beenker tiling and have demonstrated their characteristic self-similar diffraction patterns and higher-dimensional quasiperiodic structure~\cite{Viebahn2019}. Therefore, ultracold atom quasicrystals offer a highly controllable setting in which the evolution of thermodynamic, magnetic, and transport signatures of NFL behavior can be systematically investigated as functions of filling fraction, interaction strength, and quasiperiodic geometry.

Quasicrystals provide a clean platform because their critical states arise from deterministic quasiperiodic order rather than from disorder. However, similar mechanisms may operate beyond conventional quasicrystalline systems with singular or multifractal electronic states, including moir\'e van-der-Waals materials and engineered cold-atom lattices~\cite{Viebahn2019,Moon2019}. Extending the present framework to these platforms may provide a route to designing NFL behavior through geometry and spectral engineering rather than through strong correlations alone.

\section{Conclusion}
\label{sec:conclusion}
In this paper, we 
have calculated non-Fermi-liquid behaviors of various physical properties in 
weakly interacting two-dimensional quasicrystals.
In the Penrose and the Ammann--Beenker tilings, the fragmented spectrum and density of states produce anomalous temperature dependences of the Sommerfeld coefficient, magnetic susceptibility and electrical resistivity, while inhomogeneous local density of states leads to finite residual scattering rate. This not only provides a compelling theoretical explanation for experimentally observed anomalies, but also establishes a phenomenological framework capable of describing weakly interacting quasicrystals beyond the conventional Landau Fermi-liquid paradigm, with quasiperiodic criticality playing a central organizing role. The Wilson ratio further reveals that spin-fluctuation-driven non-Fermi-liquid behavior dominates over a wide range of filling fractions 
even for a weak 
interaction strength, while its magnitude and temperature dependence vary strongly with filling. 

Our works have clarified that multifractal critical electronic states in quasicrystals give rise to different types and characteristics of non-Fermi-liquid behaviors. These results establish quasicrystals as a platform for exploring various non-Fermi-liquid physics beyond the strong-correlation paradigm and open a route toward engineering anomalous metallic states through quasiperiodic geometry, filling control, and wave-function criticality. Furthermore, we have shown that the interplay between quasiperiodicity and electron-electron interactions generates an inhomogeneous temperature dependence of the Hartree shift, which reconstructs the low-energy electronic structure and further enriches the resulting non-Fermi-liquid behavior.
Our results also provide a useful guideline for analyzing experimental data measured for quasicrystals.

\begin{acknowledgments}
 We thank Kazuhiko Deguchi for valuable discussions.
 This work was supported by JSPS KAKENHI Grant No.~JP25H01397, JP25H01398, JP25K24854, and JP25K24855.
\end{acknowledgments}

\bibliography{reference_NFL}

\section*{Appendix}
\renewcommand{\thefigure}{A\arabic{figure}}
\setcounter{figure}{0}
\renewcommand{\theequation}{A\arabic{equation}}
\setcounter{equation}{0}
\subsection{Perpendicular space of quasicrystals}
\label{A1}
\subsubsection{Cut-and-project construction}
Canonical quasicrystals such as the Penrose and Ammann--Beenker tilings
can be generated by the cut-and-project construction~\cite{Duneau1985,baake2013aperiodic}.
One starts from a hypercubic lattice
$\mathbb Z^D$
embedded in a higher-dimensional Euclidean space
$\mathbb R^D$,
which is decomposed into the direct sum

\begin{equation}
\mathbb R^D
=
E_{\parallel}
\oplus
E_{\perp},
\end{equation}
where
$E_{\parallel}$
is the physical space and
$E_{\perp}$
is the perpendicular (internal) space. Quasicrystal requires irrational angle between hypercubic lattice plane and $E_{\parallel}$.

A lattice point
$\mathbf n\in\mathbb Z^D$
is projected onto the physical space by
\begin{equation}
\mathbf r
=
\Pi_{\parallel}
\mathbf n,
\end{equation}
while its internal coordinate is
\begin{equation}
\mathbf r_{\perp}
=
\Pi_{\perp}
\mathbf n.
\end{equation}
Here, $\Pi_{\parallel}$ and $\Pi_{\perp}$ are projection maps onto physical and perpendicular space, respectively. Note that due to the incommensurate angle between hypercubic lattice plane and physical space, one should retain only lattice points whose perpendicular coordinates lie inside a bounded
acceptance window
$W\subset E_{\perp}$, i.e.,
\begin{equation}
\mathbf r_{\perp}\in W.
\end{equation}
The resulting projected point set forms a quasiperiodic tiling in the
physical space.

For example, the Penrose tiling is obtained from
$\mathbb Z^5$
with
$\dim E_{\parallel}=2$
and
$\dim E_{\perp}=3$,
whereas the Ammann--Beenker tiling is generated from
$\mathbb Z^4$
with
$\dim E_{\parallel}=2$
and
$\dim E_{\perp}=2$.

\subsubsection{Reconstruction of the higher-dimensional lift}
Due to the incommensurate angle between hypercubic lattice plane and physical space, the quasicrystalline lattice points on the physical space can be lifted to the hypercubic lattice. Moreover, the projection images on physical and perpendicular spaces admit 
one-to-one correspondence.
Below, we explain this higher-dimensional lift of quasicrystalline lattice point to reconstruct perpendicular space profile. For concrete argument, we 
take the Penrose tiling as an example, however, it applies, in the same form, to various quasicrystals generated
by a cut-and-project construction, such as the Ammann--Beenker tiling. The only required ingredients are the physical
coordinates of the vertices, the nearest-neighbor graph (comprised of edges with hopping magnitude $t$), and the
correspondence between physical bond directions and the canonical basis
vectors of the high-dimensional embedding lattice. We describe the construction explicitly
for the Penrose tiling, whose natural embedding lattice is
$\mathbb Z^5$.

Let the five physical-space directions be
\begin{equation}
\mathbf e_j^{\parallel}
=
\left(
\cos\frac{2\pi j}{5},
\sin\frac{2\pi j}{5}
\right),
\qquad
j=0,\ldots,4 .
\end{equation}
A point
\begin{equation}
\mathbf n=(n_0,n_1,n_2,n_3,n_4)\in\mathbb Z^5
\end{equation}
is projected onto physical space as
\begin{equation}
\Pi_{\parallel}\mathbf n
=
\sum_{j=0}^{4}
n_j\mathbf e_j^{\parallel}.
\end{equation}
Let
$\mathbf r_i=(x_i,y_i)$ be a given projected position in the physical space. In order to compute the corresponding
perpendicular-space coordinates, we reconstruct an integer lift
\begin{equation}
i\mapsto \mathbf n_i\in\mathbb Z^5
\end{equation}
for all vertices of the finite tiling patch.


\paragraph{Edge labels as a $\mathbb Z^5$-valued one-form.}

Let $G=(V_G,E_G)$ be the nearest-neighbor graph of the tiling patch, where
$V_G$ is the set of vertices and $E_G$ is the set of oriented nearest-neighbor
bonds. For an oriented edge $(i,k)\in E_G$, the displacement
$\mathbf r_k-\mathbf r_i$ is parallel to one of the five Penrose bond
directions. Hence there exists a unique pair $(j,\sigma)$, with
$j\in\{0,\ldots,4\}$ and $\sigma=\pm1$, such that
\begin{equation}
\mathbf r_k-\mathbf r_i
=
\sigma\mathbf e_j^{\parallel}.
\end{equation}
We assign to this oriented edge the $\mathbb Z^5$-valued label
\begin{equation}
\omega(i,k)
=
\sigma\mathbf E_j,
\end{equation}
where
\begin{equation}
(\mathbf E_j)_i
=\delta_{ij}
\end{equation}
is the $j$-th canonical basis vector of $\mathbb Z^5$. Note that the label satisfies
\begin{equation}
\omega(k,i)=-\omega(i,k).
\end{equation}
Thus, $\omega$ is a discrete $\mathbb Z^5$-valued one-form on the
nearest-neighbor graph.

The desired lift is a function
\begin{equation}
\mathbf n:V_G\to\mathbb Z^5
\end{equation}
such that, for every oriented edge $(i,k)$,
\begin{equation}
\mathbf n_k-\mathbf n_i
=
\omega(i,k).
\label{eq:lift_edge_condition}
\end{equation}
In other words, $\mathbf n$ is a discrete potential whose gradient is
$\omega$.


\paragraph{Existence and path independence.}

Choose an arbitrary reference vertex $i_0\in V_G$ and fix
\begin{equation}
\mathbf n_{i_0}=\mathbf 0.
\end{equation}
For any vertex $i\in V_G$, choose a path
\begin{equation}
\gamma_G:i_0=i_0,i_1,\ldots,i_m=i
\end{equation}
from $i_0$ to $i$ and define
\begin{equation}
\mathbf n_i^{(\gamma_G)}
=
\sum_{\ell=0}^{m-1}
\omega(i_\ell,i_{\ell+1}).
\label{eq:path_integral_lift}
\end{equation}
The lift is well defined if this expression is independent of the chosen
path.

Let $\gamma_G$ and $\gamma'_G$ be two paths from $i_0$ to $i$. Then
$\gamma_G-\gamma'_G$ forms a closed cycle $\Gamma_G$. The two path integrals
give the same result if and only if
\begin{equation}
\sum_{(a,b)\in\Gamma_G}
\omega(a,b)
=
\mathbf0.
\label{eq:closed_loop_condition}
\end{equation}
Hence, the lift is well defined when the above equation holds
for every closed cycle $\Gamma_G$ in the graph.

We now prove that Eq.~\eqref{eq:closed_loop_condition} holds for a
Penrose tiling (and any quasicrystals constructed by the canonical cut-and-project scheme). Suppose that the vertices of the tiling are projections of lattice
points
\begin{equation}
\widetilde{\mathbf n}_i\in\mathbb Z^5,
\qquad
\mathbf r_i=\Pi_\parallel\widetilde{\mathbf n}_i .
\end{equation}
For a nearest-neighbor edge $(i,k)$, by construction there exists
$j$ and $\sigma=\pm1$ such that
\begin{equation}
\widetilde{\mathbf n}_k-\widetilde{\mathbf n}_i
=
\sigma\mathbf E_j .
\end{equation}
Applying $\Pi_\parallel$ gives
\begin{equation}
\mathbf r_k-\mathbf r_i
=
\Pi_\parallel
\left(
\widetilde{\mathbf n}_k-\widetilde{\mathbf n}_i
\right)
=
\sigma
\Pi_\parallel\mathbf E_j
=
\sigma\mathbf e_j^\parallel .
\end{equation}
Therefore, the edge label defined from the physical displacement satisfies
\begin{equation}
\omega(i,k)
=
\widetilde{\mathbf n}_k-\widetilde{\mathbf n}_i .
\label{eq:omega_exact}
\end{equation}

Now consider any closed cycle
\begin{equation}
\Gamma_G:
i_0,i_1,\ldots,i_m=i_0 .
\end{equation}
Using Eq.~\eqref{eq:omega_exact}, we obtain
\begin{align}
\sum_{\ell=0}^{m-1}
\omega(i_\ell,i_{\ell+1})
&=
\sum_{\ell=0}^{m-1}
\left(
\widetilde{\mathbf n}_{i_{\ell+1}}
-
\widetilde{\mathbf n}_{i_\ell}
\right)
\\ \nonumber
&=
\left(
\widetilde{\mathbf n}_{i_1}-\widetilde{\mathbf n}_{i_0}
\right)
+
\left(
\widetilde{\mathbf n}_{i_2}-\widetilde{\mathbf n}_{i_1}
\right)\nonumber\\
&\quad +\cdots
+
\left(
\widetilde{\mathbf n}_{i_0}-\widetilde{\mathbf n}_{i_{m-1}}
\right)
\nonumber \\ 
&=
\mathbf0 \nonumber.
\end{align}
Thus the closed-loop condition
\begin{equation}
\sum_{\Gamma}\omega=\mathbf0
\end{equation}
is satisfied. Consequently, the path integral
Eq.~\eqref{eq:path_integral_lift} is independent of the chosen path, and
the reconstructed lift $\mathbf n_i$ is well defined.

This also proves uniqueness up to a global translation. If
$\mathbf n_i$ and $\mathbf n'_i$ are two solutions of
Eq.~\eqref{eq:lift_edge_condition}, then
\begin{align}
(\mathbf n'_k-\mathbf n_k)
-
(\mathbf n'_i-\mathbf n_i)
&=
\left(
\mathbf n'_k-\mathbf n'_i
\right)
-
\left(
\mathbf n_k-\mathbf n_i
\right)\nonumber\\
&=
\omega(i,k)-\omega(i,k)
=
\mathbf0
\end{align}
for every edge $(i,k)$. Since the graph is connected,
\begin{equation}
\mathbf n'_i-\mathbf n_i
=
\mathbf C
\end{equation}
is independent of $i$. Hence
\begin{equation}
\mathbf n'_i
=
\mathbf n_i+\mathbf C,
\qquad
\mathbf C\in\mathbb Z^5 .
\end{equation}
The lift is therefore unique up to the arbitrary choice of the origin in
the embedding lattice.


\paragraph{Reconstruction}
Now starting from a reference site $i_0$ with
$\mathbf n_{i_0}=\mathbf0$, one visits neighboring vertices
sequentially. If the lift of vertex $i$ is already known and $k$ is an
unvisited neighbor satisfying
\begin{equation}
\mathbf r_k-\mathbf r_i
=
\sigma\mathbf e_j^\parallel ,
\end{equation}
then one assigns
\begin{equation}
\mathbf n_k
=
\mathbf n_i+\sigma\mathbf E_j .
\end{equation}
Because the closed-loop condition has been proven above, this assignment
does not depend on which path first reaches the vertex $k$. Therefore, this provides the same lift as the path integral of
Eq.~\eqref{eq:path_integral_lift}.


\paragraph{Projection to perpendicular space.}

After the integer lift has been reconstructed, the perpendicular-space
coordinate is obtained by applying the complementary projection
$\Pi_\perp$. For the Penrose tiling~\cite{deBruijn1981}, we use the internal basis
\begin{align}
x_{\perp,i}
&=
\sum_{j=0}^{4}
n_{i,j}
\cos\frac{4\pi j}{5},
\\
y_{\perp,i}
&=
\sum_{j=0}^{4}
n_{i,j}
\sin\frac{4\pi j}{5},
\\
z_{\perp,i}
&=
\frac{1}{\sqrt5}
\sum_{j=0}^{4}
n_{i,j}.
\end{align}
Thus
\begin{equation}
\mathbf r_{\perp,i}
=
(x_{\perp,i},y_{\perp,i},z_{\perp,i})
\end{equation}
is the reconstructed 
perpendicular-space coordinate of vertex $i$.

The global ambiguity
$\mathbf n_i\mapsto\mathbf n_i+\mathbf C$ produces only a uniform shift
of all perpendicular-space coordinates,
\begin{equation}
\mathbf r_{\perp,i}
\mapsto
\mathbf r_{\perp,i}
+
\Pi_\perp\mathbf C ,
\end{equation}
and therefore does not affect the relative distribution of vertices in
perpendicular space.

The above construction is independent of the specific quasiperiodic
tiling and applies to any cut-and-project quasicrystal for which the
correspondence between physical bond directions and the canonical basis
vectors of the embedding lattice is known. For example, the Ammann--Beenker tiling is generated from the projection
of the four-dimensional hypercubic lattice $\mathbb Z^4$ onto a
2D physical space.
The reconstruction algorithm remains unchanged: one assigns a
$\mathbb Z^4$-valued edge label according to the four physical bond
directions, reconstructs the integer lift by propagating these labels
along the nearest-neighbor graph, and subsequently projects the recovered
lift onto the 2D perpendicular space.
Thus, only the embedding dimension and the corresponding projection
operators are modified, while the mathematical framework and the reconstruction algorithm remain identical.

\paragraph{Perpendicular space profile}
Remind that the perpendicular-space profile provides very useful information for quasicrystalline structures and physical quantities distributed on such structures. Within the cut-and-project construction, lattice points that are close to each other in the perpendicular space generally share similar local atomic environments in the physical space. More precisely, the position of a lattice point inside the acceptance window serves as a continuous perpendicular-space coordinate that parametrizes its local environment in physical space: the closer two points are in the perpendicular space, the larger the spatial region over which their surrounding tiling patterns are locally isomorphic. Consequently, the perpendicular space offers a natural framework for classifying quasicrystalline lattice points according to their local environments, enabling the systematic analysis of local physical quantities beyond simple geometrical descriptors such as the coordination number.

\begin{figure}[h]
	\centering
	\includegraphics[width=0.5\textwidth]{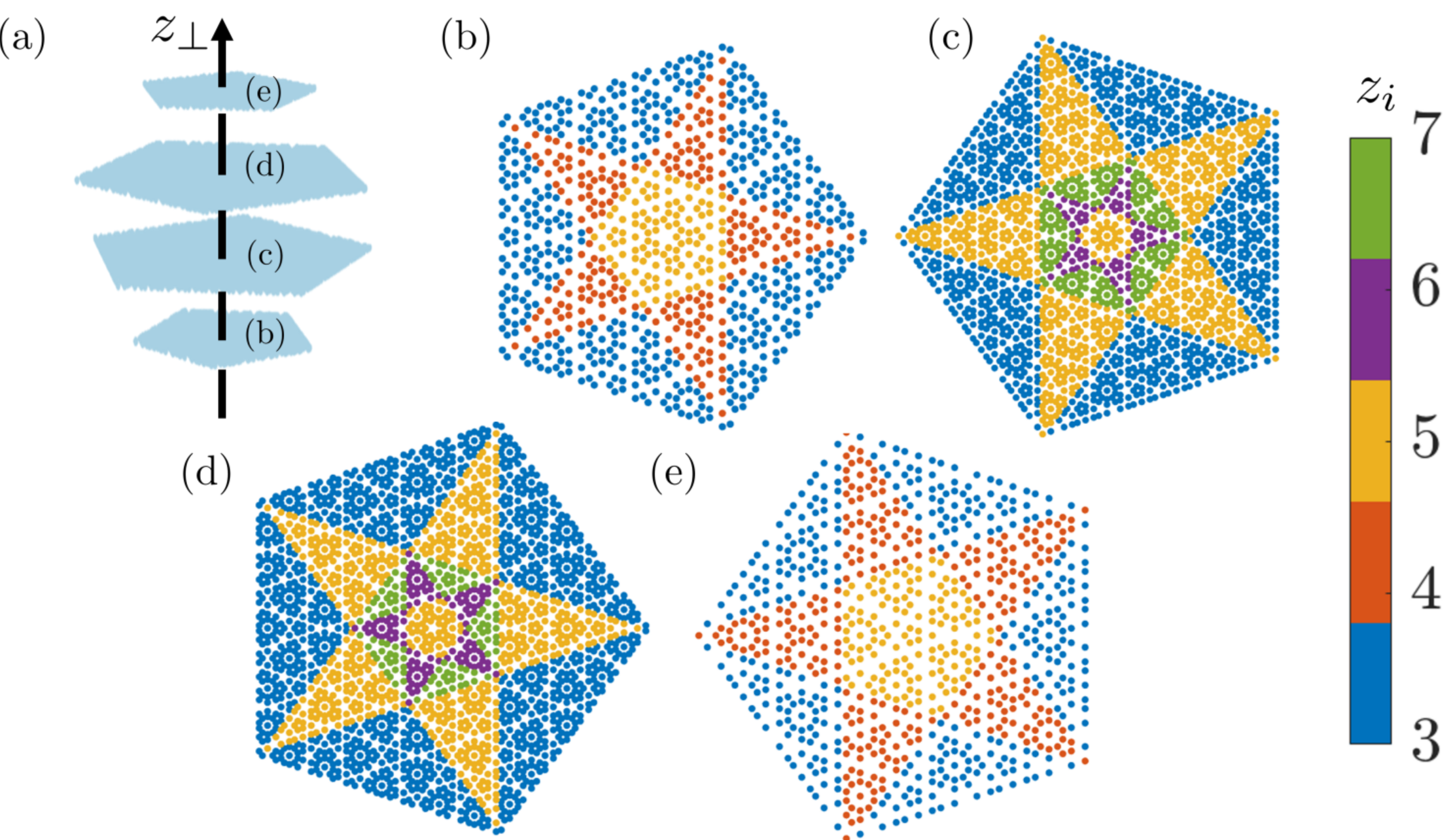}
	\caption{Perpendicular space representation of the Penrose tiling. (a) The four pentagonal acceptance windows with different $z_\perp$ corresponding to the four sublattices of the Penrose tiling. (b-e) Distribution of lattice sites classified by their coordination number $z_i$, defined as the number of nearest-neighbor bonds with hopping amplitude, $t$. Sites with 
    the same $z_i$ 
    occupy 
    the same domain,
    forming nested star-shaped 
    pattern
    that reflects the hierarchical local environments of the Penrose tiling. 
    }
	\label{fig:perpendicular_penrose}
	
\end{figure}
Figure~\ref{fig:perpendicular_penrose} shows the perpendicular-space representation of the Penrose tiling, where each lattice site is colored according to its coordination number, $z_i$. Here, $z_i$ is defined as the number of nearest-neighbor bonds with hopping amplitude $t$. The perpendicular space of the Penrose tiling consists of four pentagonal acceptance windows corresponding to the four cosets of the Penrose tiling. Within each acceptance window, lattice sites are not distributed randomly but form highly organized patterns reflecting the hierarchical and self-similar structure of the tiling. Furthermore, the coordination number is not uniformly distributed over the acceptance window. Instead, sites with the same coordination number occupy certain regions bounded by simple polygonal curves, giving rise to nested star-shaped domains [see Figs.~\ref{fig:perpendicular_penrose}(b--e)]. In particular, higher-coordination sites are concentrated near the centers of the acceptance windows, whereas lower-coordination sites predominantly appear toward their boundaries. This spatial organization demonstrates that the perpendicular-space coordinate serves as a natural descriptor of the local geometrical environment, with nearby points in perpendicular space corresponding to lattice sites that share increasingly similar local atomic configurations in the physical space.

Comparing Figures~\ref{fig:perpspace} and \ref{fig:perpendicular_penrose}, we note that not only LDOS but also residual scattering rate and local scaling exponent of self-energy show finer structures in the perpendicular space compared to the coordination number. Thus, local NFL indicators in quasicrystal feature long-range correlations.

\subsection{Thermodynamic property and magnetic response for different system sizes}
\label{A2}
\begin{figure}[h]
	\centering
	\includegraphics[width=0.5\textwidth]{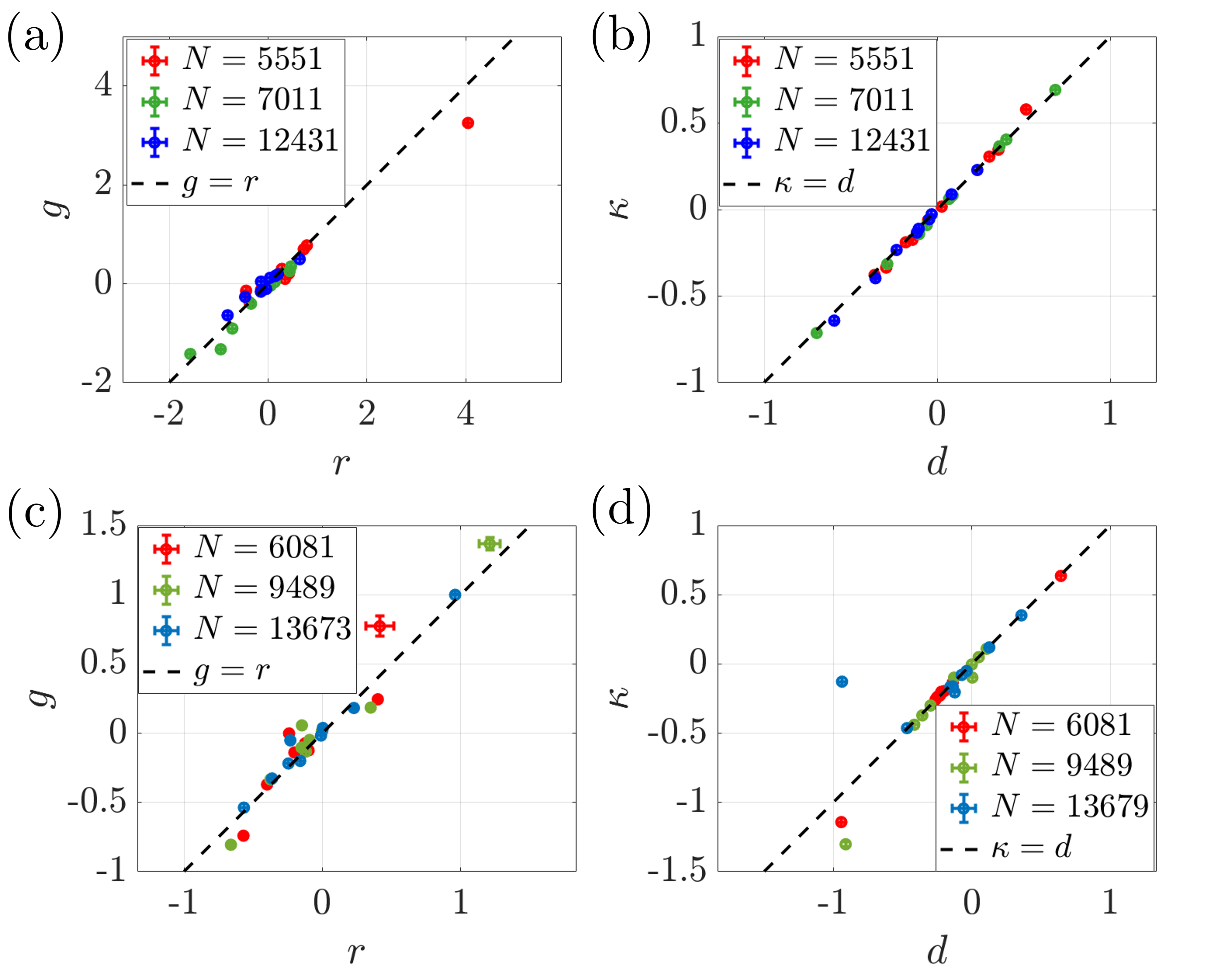}
	\caption{Finite-size dependence of the universal scaling relations between the thermodynamic quantities and the DOS for $U=0.1t$. (a,b) Penrose tilings with $N=5551$, $7011$, and $12431$. (c,d) Ammann--Beenker tilings with $N=6081$, $9489$, and $13673$. Panels (a) and (c) compare the scaling exponents $g$ and $r$, while panels (b) and (d) compare $\kappa$ and $d$. Although the scaling exponents at a fixed filling fraction change with system size due to the increasingly fragmented DOS, the universal relationships $g\simeq r$ and $\kappa \simeq d$ remain robust. The larger deviation for the $10\%$-filled Ammann--Beenker tiling originates from the enhanced singularity of DOS in larger systems.}
	\label{fig:finite_size}
	
\end{figure}

Figure~\ref{fig:finite_size} demonstrates that, although the scaling exponents at a fixed filling fraction vary with system size due to the increasingly fragmented DOS, the universal relationships, $g=r$ and $\kappa=d$, 
still hold across different system sizes. Note that the irregularity of the relationship between $\kappa$ and $d$ associated with the confined localized states in the Ammann--Beenker tiling becomes more pronounced as the system size increases, reflecting the divergence of DOS in the thermodynamic limit [see Fig.~\ref{fig:finite_size}(d)].

\subsection{Inhomogeneous temperature-dependent Hartree shift}
\label{A_add}
\begin{figure}[h]
	\centering
	\includegraphics[width=0.5\textwidth]{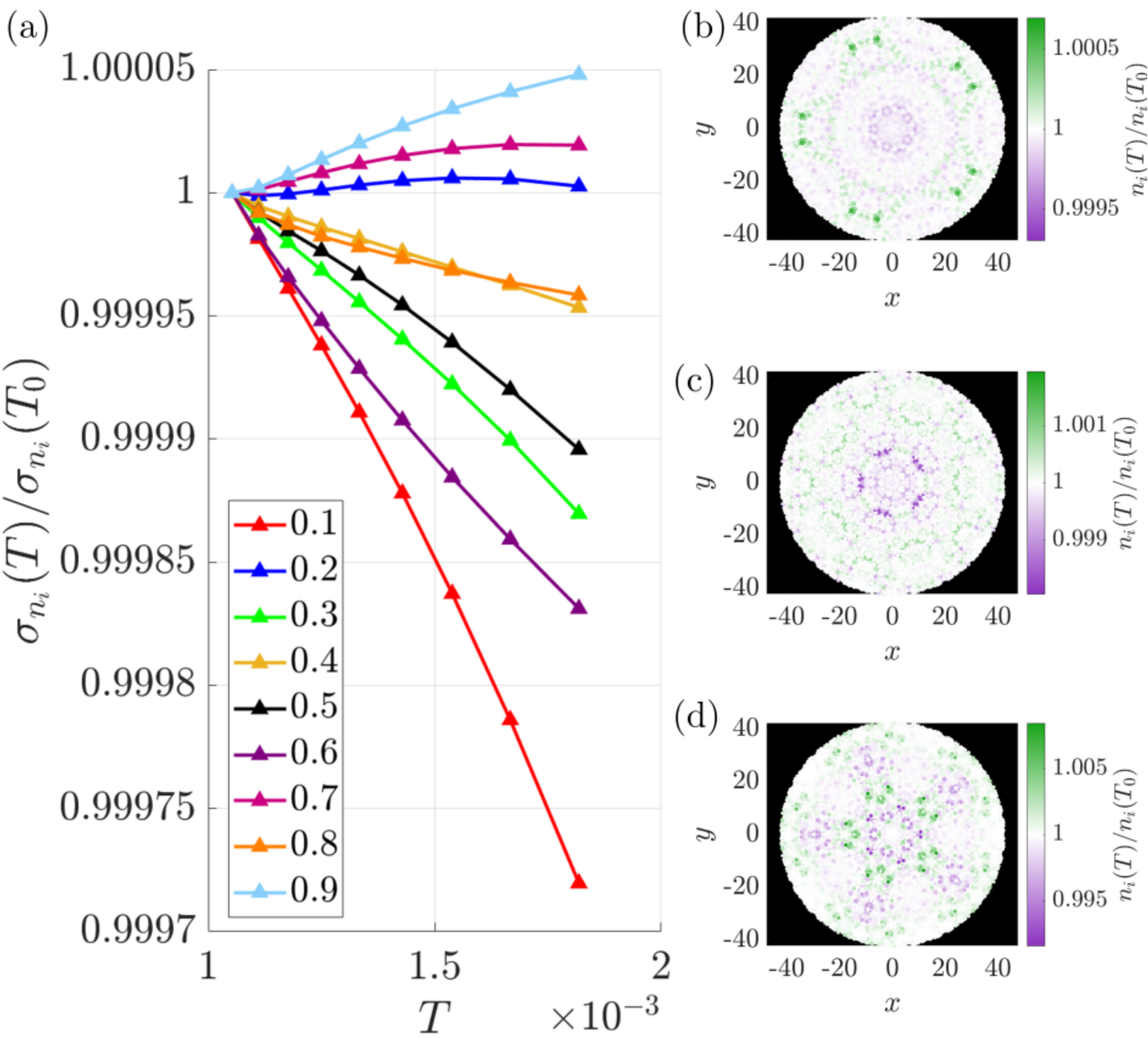}
	\caption{(a) Anomalous temperature dependence of the inhomogeneous Hartree shift on the Penrose tiling at $U=0.5t$. Standard deviation of the local charge density, $n_i$, is plotted as a function of temperature for various filling fractions. Here, the reference temperature is $T_0=t/950$. Inhomogeneous temperature dependence of $n_i$ for different filling fractions, (b) 0.9 (c) 0.4 and (d) 0.1. Here, $T=t/550$.}
	\label{fig:Hartree_inhomo_Penrose}
	
\end{figure}

\begin{figure}[h]
	\centering
	\includegraphics[width=0.5\textwidth]{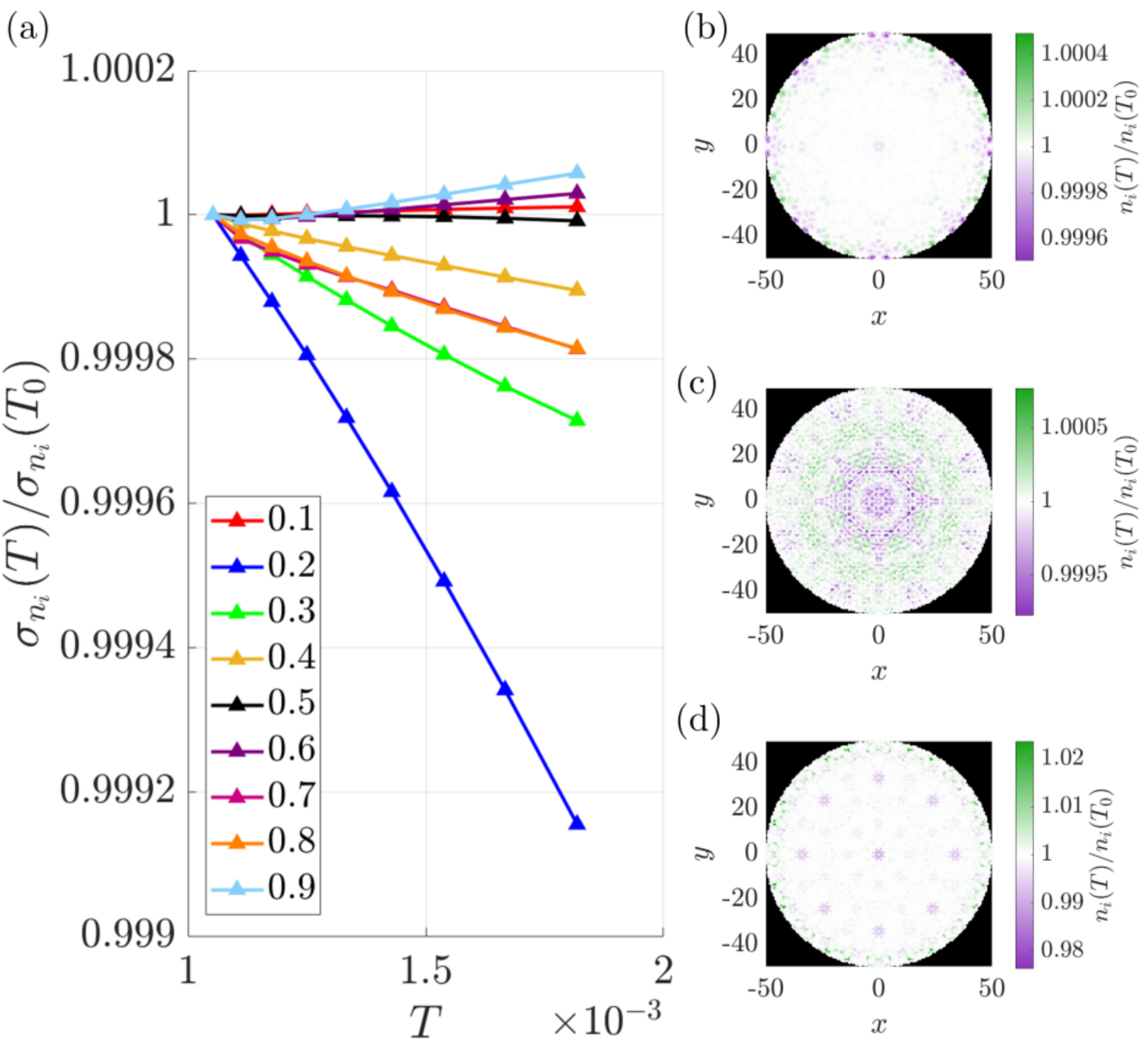}
	\caption{(a) 
    The same as Fig.~\ref{fig:Hartree_inhomo_Penrose}(a) but for the Ammann--Beenker tiling.
    Inhomogeneous temperature dependence of $n_i$ for different filling fractions, (b) 0.9 (c) 0.4 and (d) 0.2. Here, $T=t/550$.}
	\label{fig:Hartree_inhomo_AB}
	
\end{figure}

The charge density in quasicrystals generally exhibits an inhomogeneous spatial distribution (Fig.~\ref{fig:criticalHartree}), 
which depends on temperature.
This gives rise to unconventional Hartree-shift effects, which modify the DOS and 
lead to filling-dependent anomalous temperature dependences of thermodynamic quantities. 
Here, we discuss the dependence of the Hartree shift 
on the filling fraction and temperature. Figures~\ref{fig:Hartree_inhomo_Penrose} and ~\ref{fig:Hartree_inhomo_AB} exhibit 
the filling and temperature dependence of the  inhomogeneous charge distribution 
on the Penrose and Ammann-Beenker tilings, respectively.
In detail, Figures~\ref{fig:Hartree_inhomo_Penrose}(a) and \ref{fig:Hartree_inhomo_AB}(a) demonstrate that the spatial fluctuation of charge density, measured by its standard deviation $\sigma_{n_i}$, varies with temperature. 
The temperature dependence is not monotonic but strongly filling-dependent. 
For most filling fractions,
the inhomogeneity ($\sigma_{n_i}$) of charge density decreases as the temperature increases. This is because thermal average reduce the spatial fluctuation at higher temperatures. Notably, however, for some filling fractions, the inhomogeneity increases with temperature. Thus, the evolution of the inhomogeneity of the Hartree shift is strongly filling-dependent in quasicrystals. Furthermore, for general filling fractions, the temperature dependence of the local charge density $n_i$ is strongly site-dependent [see panels (b--d) in Figs.~\ref{fig:Hartree_inhomo_Penrose} and \ref{fig:Hartree_inhomo_AB}]. Note that although the charge density $n_i$ itself is not fractal, its temperature-induced variation would exhibit a fractal spatial pattern because it is governed by the LDOS near the Fermi level. Hence, the inhomogeneity of the Hartree shift plays a complicated role in reconstructing the thermal DOS as the temperature varies. This contributes to the anomalous temperature dependence of thermodynamic quantities in weakly-interacting quasicrystals.

\subsection{Inhomogeneous magnetization}
\label{A3}
\begin{figure}[h]
	\centering
	\includegraphics[width=0.5\textwidth]{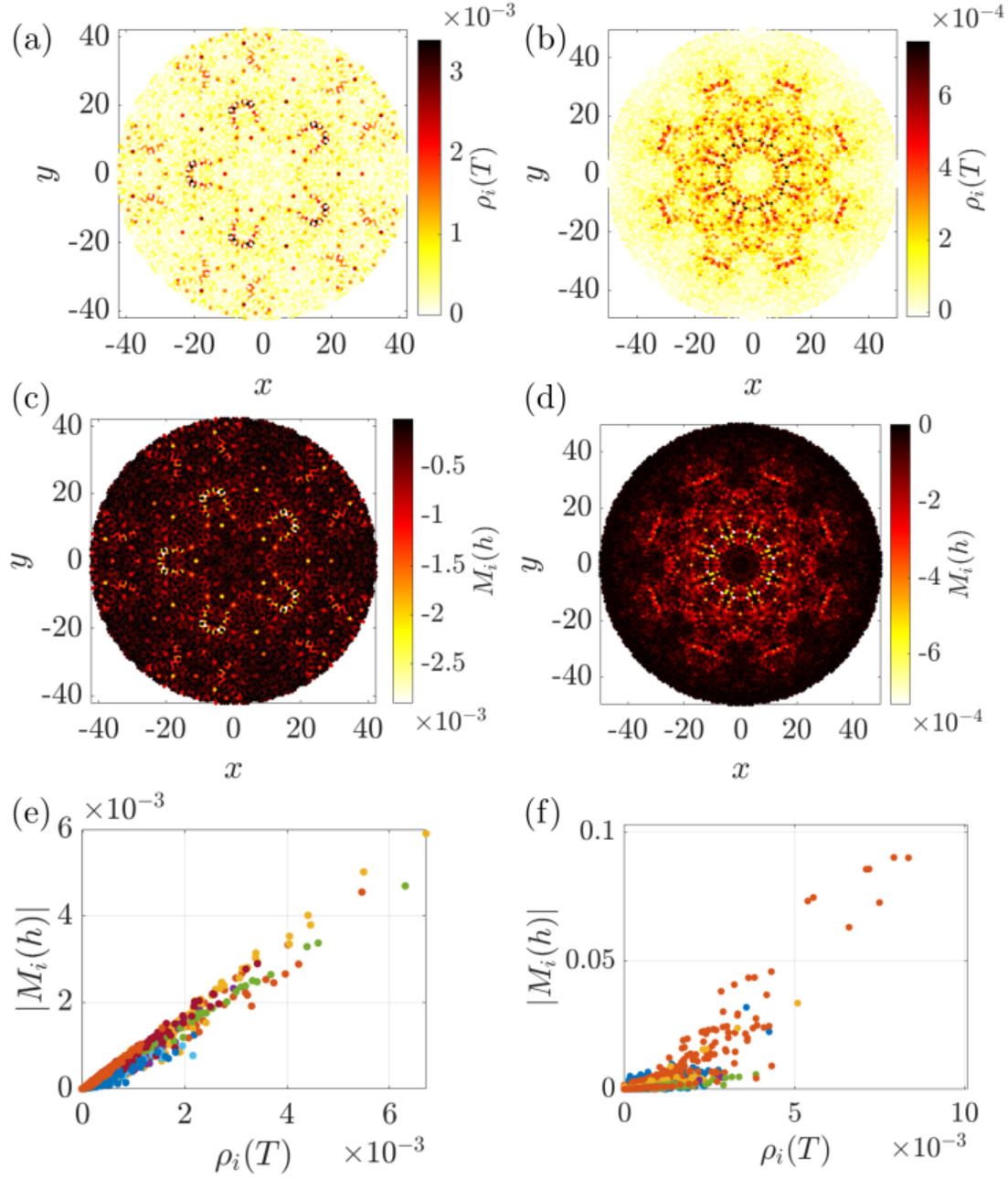}
	\caption{Real-space maps of (a,b) paramagnetic thermal LDOS, $\rho_i(T)$, and (c,d) local magnetization, $M_i(h)$, at $70\%$ filling. (e,f) 
    The magnitude of $M_i(h)$ plotted against $\rho_i(T)$. Different colors represent different filling fractions. Panels (a,c,e) [(b,d,f)] show the results for the Penrose (Ammann--Beenker) tiling.
    External magnetic field strength of  $h=10^{-3}t$, $U=0.5t$, and the temperature of $T=10^{-3}$t are used. 
    }
	\label{fig:magnetization}
	
\end{figure}

The paramagnetic response emergent in quasicrystals is spatially inhomogeneous associated with the inhomogeneous LDOS of critical states. 
Figures~\ref{fig:magnetization}(a,b) display inhomogeneous thermal LDOS, $\rho_i(T)$ in quasicrystals, while Figs.~\ref{fig:magnetization}(c,d) show inhomogeneous local magnetizations, $M_i(h)$. Note that the spatial patterns of $M_i(h)$ closely resembles that of $\rho_i(T)$. Specifically, the local magnetization is enhanced at sites with a large LDOS [see Figs.~\ref{fig:magnetization}(e,f)].

\subsection{Magnetic ordered states in quasicrystals}
\label{A4}
\begin{figure}[!h]
	\centering
	\includegraphics[width=0.5\textwidth]{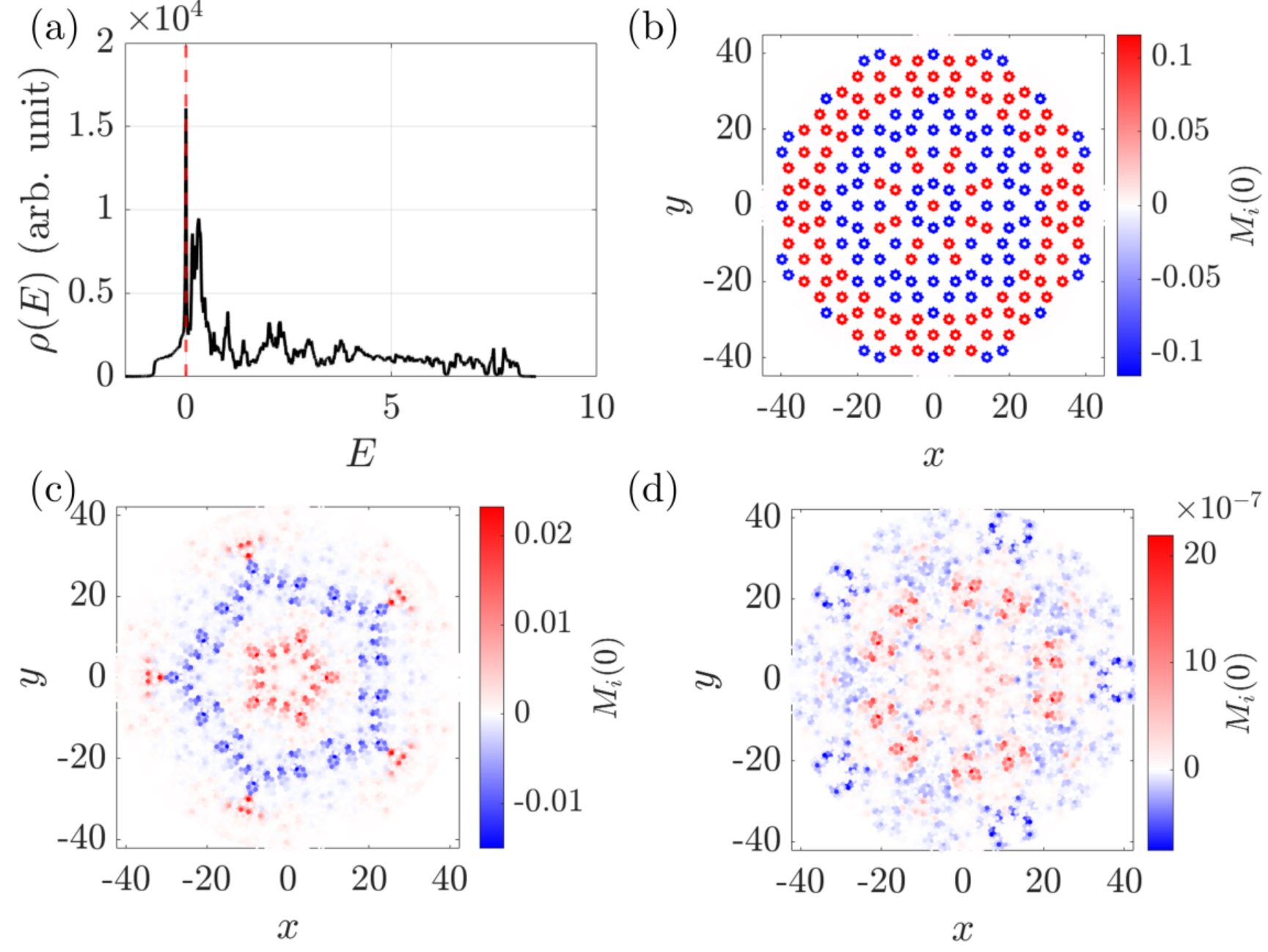}
	\caption{(a) DOS of Ammann--Beenker tiling at $10\%$ filling. Red dashed vertical line is drawn for emphasizing singularity near the Fermi energy. (b) Magnetically ordered state of $10\%$-filled Ammann--Beenker tiling for $U=0.1t$. Magnetic structure of $10\%$-filled Penrose tiling for (c) $U=t$ and (d) $U=0.5t$. Panel (c) exhibits an antiferromagnetic state, while panel (d) shows that a paramagnetic state is stabilized as local magnetization vanishes. 
    }
	\label{fig:magneticorder}
\end{figure}

Here, we explore emergent magnetic ordered states in quasicrystals. First, we focus on the \(10\%\)-filled Ammann--Beenker tiling that 
exhibits a large deviation 
from the linear scaling relation, $\kappa\simeq d$, 
even for small $U/t=0.1$ [see Figs.~\ref{fig:suscep}(d) and \ref{fig:finite_size}(d)]. This filling fraction corresponds to the flat band comprised of confined localized states [see Fig.~\ref{fig:magneticorder}(a)]. Figure~\ref{fig:magneticorder}(b) exhibits the magnetically ordered state even at small $U/t=0.1$. Note that the correspondence between the DOS and magnetic susceptibility breaks down since the random phase approximation we used is no longer valid. Hence, the two exponents, $\kappa$ and $d$, deviate from the linear relation even in the weak-interaction regime at this special filling fraction.

Next, let us consider filling fractions with a 
normal
DOS, excluding those associated with the singularity. Note that the magnetically ordered state appears at the singularity of DOS even for weakly interacting regime. In contrast, for general filling fractions, the paramagnetic state is stabilized for small $U/t$, while 
a larger $U/t$ stabilizes
magnetically ordered states. For instance, Figs.~\ref{fig:magneticorder}(c,d) exhibit the magnetic structure of $10\%$-filled Penrose tiling for $U=t$ and $U=0.5t$, respectively. For $U=t$, the intermediate strength of interaction, the antiferromagnetic order would be stabilized for some filling fractions. However, unlike the case of confined localized states, the paramagnetic soultion is stabilized for small $U/t$ as local magnetization vanishes at zero external field [see Fig.~\ref{fig:magneticorder}(d)].

\subsection{Scattering rate of Ammann-Beenker tiling}
\label{A5}
\begin{figure}[h]
	\centering
	\includegraphics[width=0.5\textwidth]{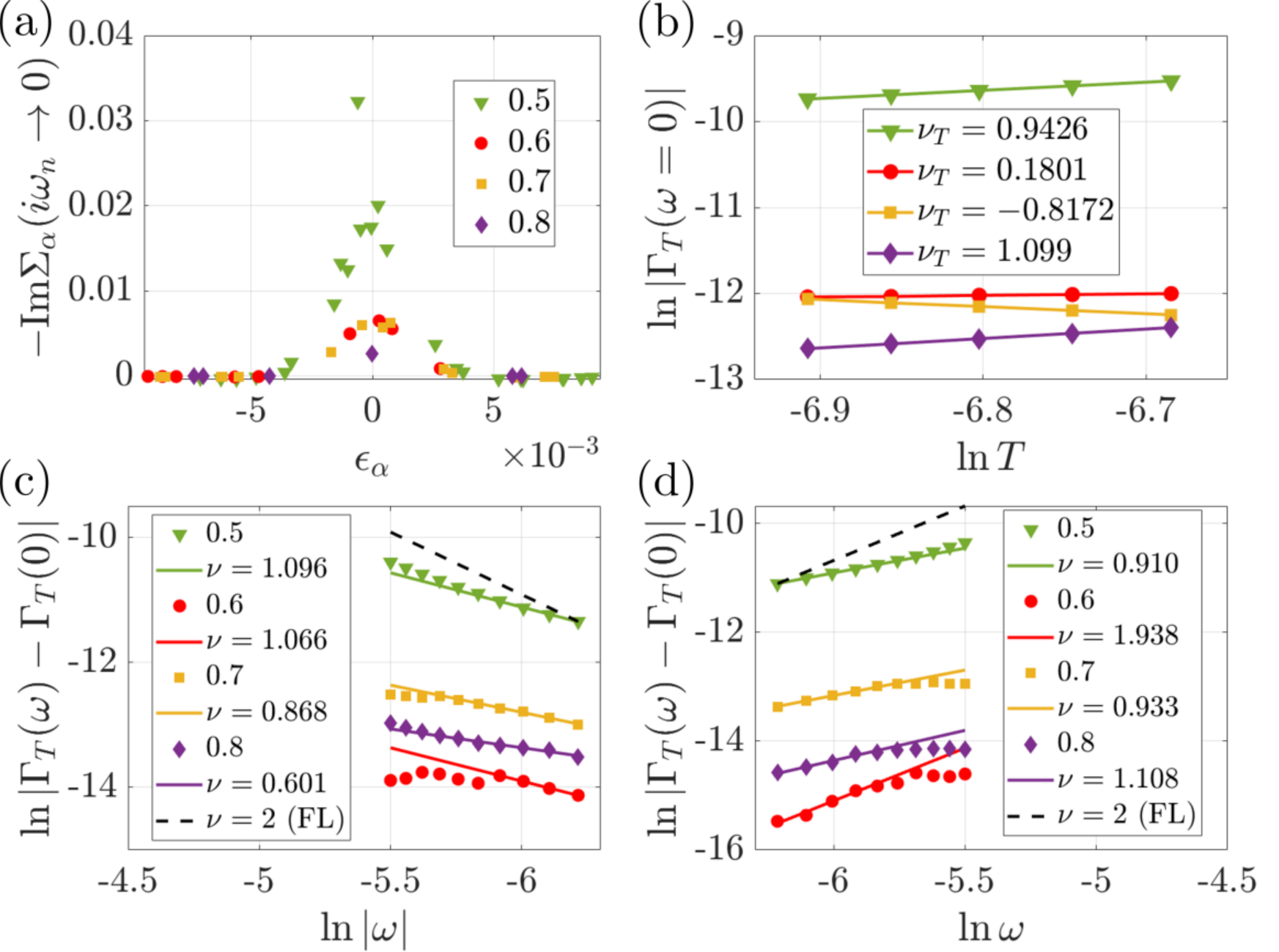}
	\caption{NFL behaviors of thermal scattering rate of Ammann-Beenker tiling for $U/t=0.5$. (a) Residual scattering rate for the quasiparticle state, $\psi_\alpha$ as a function of energy, $\epsilon_{\alpha}$. $-\mathrm{Im}\Sigma_\alpha(\mathrm{i}\omega_n\to0)>0$ suggests nonzero residual resistivity. (b) Log-log plot of zero-frequency scattering rate as a function of temperature. Green, red, gold and violet colors represent filling fractions, 0.5, 0.6, 0.7 and 0.8, respectively, where a paramagnetic solution is stable. (c,d) Scaling behaviors of thermal scatterng rate at $T=10^{-3}t$ as a function of frequency. Different colors represent different filling fractions. Black dashed line represents Fermi liquid scaling exponent, $\nu=2$. Panels (c) and (d) display negative and positive frequency results, respectively. 
    }
	\label{fig:transport_scaling_AB}
	
\end{figure}

Figure~\ref{fig:transport_scaling_AB} presents the scaling behavior of the thermal scattering rate, $\Gamma_T(\omega)$, for different filling fractions of the Ammann--Beenker tiling, where the paramagnetic solution remains stable at $U/t=0.5$. Figure~\ref{fig:transport_scaling_AB}(a) shows that a finite residual scattering rate emerges in the vicinity of the Fermi level. The scaling analyses in Figs.~\ref{fig:transport_scaling_AB}(b-d) further demonstrate that both the temperature exponent, $\nu_T$, and the frequency exponent, $\nu$, exhibit NFL behavior. In addition, comparing Figs.~\ref{fig:transport_scaling_AB}(c) and (d), we find that the frequency scaling exponent generally differs between positive and negative frequencies, reflecting the intrinsic particle-hole asymmetry of the scattering process. These results indicate that the finite residual scattering rate and the NFL scaling of $\Gamma_T(\omega)$ are robust features of weakly interacting quasicrystals, rather than properties specific to the Penrose tiling.

\begin{figure}[h]
	\centering
	\includegraphics[width=0.5\textwidth]{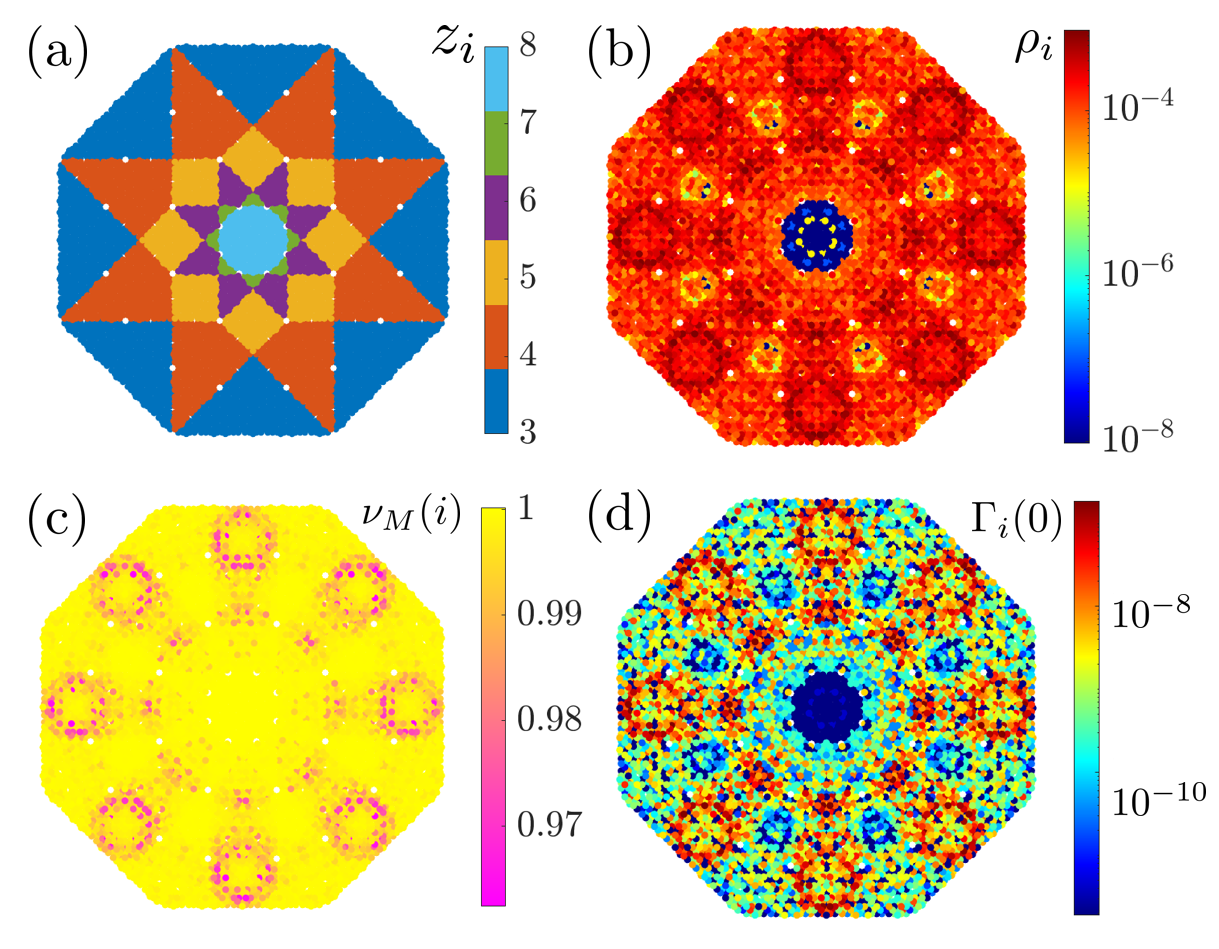}
	\caption{Perpendicular-space profile
    of the Ammann--Beenker tiling. (a) Coordination number ($z_i$), 
    (b) LDOS, $\rho_i$, (c) Matsubara scaling exponent, $\nu_M(i)$, and (d) residual scattering rate, $\Gamma_i(0)$. Unlike the coordination number, the local electronic quantities exhibit fine structures, indicating that the local NFL properties are governed by the long-range quasiperiodic environment. Here, $U=0.5t$, $T=10^{-3}t$ and the filling fraction is 0.6. 
    }
	\label{fig:transport_perp_AB}
	
\end{figure}

Similar to the Penrose tiling, the local NFL indicators are governed by long-range quasiperiodic structures rather than the coordination number. Figure~\ref{fig:transport_perp_AB}(a) shows the perpendicular space representation of the Ammann-Beenker tiling, where colors denote the coordination number of each site. Compared to this coordination-number map,
the LDOS, $\rho_i$, the local scaling exponent of Matsubara frequency self-energy, $\nu_M(i)$, and the local residual scattering rate, $\Gamma_i(0)$, all exhibit finer structures in perpendicular space [see Figs.~\ref{fig:transport_perp_AB}(b-d)]. Thus, the local NFL properties of quasicrystals are determined by the long-range environment.

\end{document}